\documentclass[prd,twocolumn,notitlepage,showpacs,preprintnumbers,amsmath,amssymb,APS,10pt,superscriptaddress,longbibliography,nofootinbib]{revtex4-2}

\pdfoutput = 1

\usepackage{dcolumn}
\usepackage{bm}
\usepackage{xcolor}
\usepackage[utf8]{inputenc}
\usepackage{amsmath,amssymb,amsfonts,latexsym,cancel}
\usepackage[normalem]{ulem}
\usepackage{graphicx}
\usepackage{color}
\usepackage{soul}
\usepackage{ulem}
\usepackage[colorlinks=true,linkcolor=red,urlcolor=blue,citecolor=blue]{hyperref}
\usepackage{amsmath}
\usepackage{slashed}
\usepackage{braket}
\usepackage{amssymb}
\usepackage{amsmath}
\usepackage{accents}
\usepackage{soul}

\usepackage{graphicx}
\allowdisplaybreaks

\newcommand{\be}{\begin{equation}}
\newcommand{\ee}{\end{equation}}
\newcommand{\bea}{\begin{eqnarray}}
\newcommand{\eea}{\end{eqnarray}}

\usepackage{orcidlink,booktabs}
\usepackage{soul,amsmath,amssymb}
\usepackage{siunitx,bm}
\usepackage{comment}

\usepackage{lettrine}
\usepackage{Zallman}

\newlength{\dhatheight}

\DeclareFontFamily{U}{wncy}{}
    \DeclareFontShape{U}{wncy}{m}{n}{<->wncyr10}{}
    \DeclareSymbolFont{mcy}{U}{wncy}{m}{n}
    \DeclareMathSymbol{\Sh}{\mathord}{mcy}{"58}

\begin{document}

\title{Ultralight dark matter detection with trapped-ion interferometry}

\author{Leonardo Badurina}
\affiliation{Walter Burke Institute for Theoretical Physics, California Institute of Technology, Pasadena, CA 91125, USA}

\author{Diego Blas}
\affiliation{Institut de F\'{i}sica d’Altes Energies (IFAE), The Barcelona Institute of Science and Technology, Campus UAB, 08193 Bellaterra (Barcelona), Spain}
\affiliation{Instituci\'{o} Catalana de Recerca i Estudis Avan\c{c}ats (ICREA),
Passeig Llu\'{i}s Companys 23, 08010 Barcelona, Spain}

\author{John Ellis}
\affiliation{Theoretical Particle Physics and Cosmology (TPPC) Group, Department of Physics, \\King's College London, Strand, London, WC2R 2LS, UK} 
\affiliation{Theoretical Physics Department, CERN, CH-1211 Geneva 23, Switzerland}

\author{Sebastian A. R. Ellis}
\affiliation{Département de Physique Théorique, Université de Genève,
24 quai Ernest Ansermet, 1211 Genève 4, Switzerland}

\date{\today}

\begin{abstract}
We explore how recent advances in the manipulation of single-ion wave packets open new avenues for detecting weak magnetic fields sourced by ultralight dark matter. 
A trapped ion in a ``Schr\"odinger cat'' state can be prepared with its spin and motional degrees of freedom entangled and be used as a matter-wave interferometer that is sensitive to the Aharonov-Bohm-like phase shift accumulated by the ion over its trajectory. The result of the spin-motion entanglement is a parametrically-enhanced sensitivity to weak magnetic fields as compared with an un-entangled ion in a trap. Taking into account the relevant boundary conditions, we demonstrate that a single trapped ion can probe unexplored regions of kinetically-mixed dark-photon dark matter parameter space in the $10^{-15}~\text{eV} \lesssim m_{A'}
\lesssim 10^{-14}$~eV mass window. We also show how such a table-top quantum device will also serve as a complementary probe of axion-like particle dark matter in the same mass window.
\end{abstract}

\preprint{CALT-TH/2025-011, KCL-PH-TH/2025-29, CERN-TH-2025-125, AION-REPORT/2025-05}

\maketitle

\section{Introduction.} 
Determining the microphysical nature of dark matter (DM) remains one of the outstanding problems in particle physics and cosmology. 
The parameter space of DM mass and coupling strengths to Standard Model (SM) particles is vast. 
Covering all possible candidates therefore requires a similarly broad experimental program.

Among the landscape of DM candidates, a possibility receiving increasing attention is that DM is that of an ultra-light boson of mass $m_{\rm DM}\lesssim 1\,$eV. 
Examples include CP-even scalars such as dilatons~\cite{Goldberger:1999uk,Low:2001bw,Damour:2010rp,Hui:2016ltb,Flacke:2016szy,Ferreira:2020fam,Banerjee:2022sqg,Hubisz:2024hyz}, CP-odd pseudoscalars such as axions and axion-like particles (ALPs)~\cite{Peccei:1977hh,Peccei:1977ur,Weinberg:1977ma,Wilczek:1977pj,Shifman:1979if,Kim:1979if,Dine:1981rt,Zhitnitsky:1980tq,Preskill:1982cy,Abbott:1982af,Dine:1982ah,Svrcek:2006yi,Arvanitaki:2009fg}, dark photons (DP)~\cite{Holdom:1985ag,Abel:2008ai,Goodsell:2009xc,Nelson:2011sf,Arias:2012az,Graham:2015rva,Agrawal:2018vin,Bastero-Gil:2018uel,Dror:2018pdh,Co:2018lka,Caputo:2021eaa,Adshead:2023qiw,Cyncynates:2023zwj,Cyncynates:2024yxm} or spin-2 massive states \cite{Blas:2025qrw}. 
These models generically feature couplings to SM photons. 
For example, an ALP $a$ may couple to a pair of electromagnetic fields via the dimension-5 operator $\mathcal{O} \sim g_{a\gamma\gamma}a F_{\mu\nu}\widetilde{F}^{\mu\nu}$, where $g_{a\gamma\gamma}$ is a dimensionful coupling. 
A dark photon $A'_\mu$ may couple to SM photons via kinetic mixing, i.e., the Lagrangian may feature the dimension-4 operator $\epsilon F_{\mu\nu}F'^{\mu\nu}$, where $\epsilon \ll 1$ is the dimensionless kinetic mixing parameter and $F'^{\mu\nu}$ is the field tensor of the dark photon. 

In light of their large occupation number per de Broglie volume, these ultra-light dark matter (ULDM) candidates can be described as classical waves, whose oscillation frequency is largely determined by the DM mass. Via the aforementioned interactions, ALP waves may convert into electromagnetic radiation in the presence of static electromagnetic fields, while ``dark'' electromagnetic fields associated with kinetically-mixed dark photons may source electromagnetic fields.

Crucially, the electromagnetic fields sourced by these candidates are small. 
To measure such small signatures over a wide range of masses, a number of laboratory concepts have been proposed. 
For recent reviews of the experimental landscape, see Refs.~\cite{Adams:2022pbo} and~\cite{Antypas:2022asj} and references therein.
Most searches for dark photon dark matter take advantage of the fact that it naturally generates an $\epsilon$-suppressed electromagnetic field that can be measured in the laboratory (see, e.g., Ref.~\cite{Caputo:2021eaa} for a discussion of these signals).
Meanwhile, searches for axion or ALP dark matter typically require an applied magnetic field to trigger axion-photon conversion~\cite{Sikivie:1983ip}.
While the basic principles underlying searches for these dark matter candidates are well-known, there has been significant recent work leveraging new experimental techniques to extend the sensitivity of detectors to a wider range of parameter space~\cite{Chaudhuri:2014dla,Kahn:2016aff,Lawson:2019brd,Berlin:2019ahk,Lasenby:2019prg,Berlin:2020vrk,Schutte-Engel:2021bqm,ALPHA:2022rxj,Marsh:2022fmo,DMRadio:2022jfv,Kuenstner:2022gyc,DMRadio:2023igr,Higgins:2023gwq,An:2023wij,Beadle:2024jlr,Kalia:2024eml,An:2024wmc,Beadle:2025dgy}.

In this work, we propose searching for kinetically-mixed dark photon dark matter and ALPs with trapped-ion interferometers. 
Our concept takes advantage of the enormous improvement in the control and development of small-scale ion traps, primarily for quantum computing, and makes use of the associated advances in manipulation of the ion's quantum states~\cite{Cats_1996,PhysRevLett.105.090502,PhysRevLett.104.140501,PhysRevLett.110.203001,APB_control_2014,Campbell_2017,West:2019xio,West:2020qns,Shinjo_2021,Putnam:2023sqt}.
Leveraging the entanglement between the ion's spin and motional degrees of freedom, trapped-ion matter-wave interferometers using ``Schr\"odinger cat" (also known as  ``cat") states~\cite{Cats_1996}, have been proposed to measure the Sagnac effect~\cite{Campbell_2017}, namely the relative phase shift between counter-propagating beams in a loop interferometer induced by the rotation of the interferometer's reference frame~\cite{CRPHYS_2014__15_10_875_0}.
Here we demonstrate that a trapped-ion matter-wave interferometer using a cat state is sensitive to the Aharonov-Bohm phase induced by weakly-coupled dark matter fields. 
Considering the relevant boundary conditions, we demonstrate that such a quantum sensor can probe unexplored regions of ALP and DP parameter space.

\section{Trapped-ion Interferometry.} 
In this work, we consider the protocol proposed in Ref.~\cite{Campbell_2017}. An ion of mass $m_{\rm ion}$ is initially cooled and prepared in a linear
superposition of internal quantum states $\ket{\uparrow}$ and $\ket{\downarrow}$
inside a harmonic and rotationally symmetric trap.
The two wavepackets are subsequently set in counter-propagating orbital motion around the trap centre following $N$ successive spin-dependent kicks (SDKs), which entangle the motional and internal degrees of freedom and separate the two states by $2 N k_{\rm eff}$ in momentum space, where $k_\mathrm{eff}$ is the momentum transferred to one wavepacket after a single SDK, and a non-adiabatic trap displacement
$y_d$ in a direction orthogonal to the SDKs. As a result of their dynamics around the trap centre, the ion is in a linear superposition of spatially delocalised coherent states (i.e., cat states). After an integer number of revolutions, corresponding to an interrogation time $\Delta t$, an interference measurement is performed in the qubit basis $\{\ket{+},\ket{-}\}$.
A complete description of the protocol is given in Appendix~\ref{app:protocol}.
In the set-up envisaged in this work, we assume that the trap confines the motion of the ion to two dimensions, within which each wavepacket encloses an area $|\boldsymbol{\Sigma}|$ after a single revolution.
Crucially, since each wavepacket has the same electric charge, the trapped ion will accumulate an Aharonov-Bohm phase shift that depends exclusively on the magnetic vector potential $\boldsymbol{A}$:
\begin{equation}\label{eq:phase_shift}
\begin{aligned}
\Delta \Phi &= 2 e \oint \boldsymbol{A}\cdot\mathrm{d}\boldsymbol{x} \, .
\end{aligned}
\end{equation}
Using Stokes' Theorem and the assumption that the vector potential is spatially uniform over the extent of the region of interest, which is a good approximation in our setup, Eq.~\eqref{eq:phase_shift} may be written the form
\begin{equation}\label{eq:phase_shift2}
\begin{aligned}
\Delta \Phi &\simeq e \int_0^{\Delta t} (\boldsymbol{B}\times \boldsymbol{x})\cdot\frac{\mathrm{d}\boldsymbol{x}}{\mathrm{d}t}\, \mathrm{d}t \\
& = 2 e \int_0^{\Delta t} \boldsymbol{B} \cdot \frac{\mathrm{d}\boldsymbol{\Sigma}}{\mathrm{d}t} \mathrm{d}t \ .
\end{aligned}
\end{equation}
The second line of Eq.~\eqref{eq:phase_shift2} follows from vector identities and the rate of change of the area subtended by the wavepackets during their orbital motion, i.e., $2 \,  \mathrm{d}\boldsymbol{\Sigma}/\mathrm{d}t \equiv \boldsymbol{x} \times \mathrm{d}\boldsymbol{x}/\mathrm{d}t$. In terms of experimentally tunable parameters, $2 \,  \mathrm{d}\boldsymbol{\Sigma}/\mathrm{d}t = y_d N k_{\rm eff}/m_{\rm ion} \hat{\boldsymbol{\Sigma}}$, where $\hat{\boldsymbol{\Sigma}}$ is a unit vector perpendicular to the area enclosed by the wavepackets. While Eq.~\eqref{eq:phase_shift} is written in the usual form of a true Aharonov-Bohm effect, it should be noted that the ion evolves in a region where the magnetic field $\boldsymbol{B}$ is non-zero. Therefore, while the phase shift can be thought of as Aharonov-Bohm-like, it is not strictly speaking a topological phase.

The second line of Eq.~\eqref{eq:phase_shift2} may equivalently be interpreted as the ion's \textit{dynamical Zeeman phase shift} in the presence of a background magnetic field, where the effective magnetic moment of the ion is defined as $\mu = 2N k_{\rm eff}\, y_d(m_e/m_\mathrm{ion})\mu_B$, where $m_e$ is the electron mass and $\mu_B$ is the Bohr magneton. For the parameters used in our projections, 
$\mu/\mu_\mathrm{B} \approx 1$ so that, after an interrogation time $\Delta t = 1$~s, a detector employing a single ion limited by shot noise (i.e., with phase shift sensitivity $1~\mathrm{rad}/\sqrt{\mathrm{Hz}}$) would have a sensitivity to a magnetic field of approximately $5\times10^{-12}~\mathrm{T}/\sqrt{\mathrm{Hz}}$. 
We note that the noise level is significantly greater than that of other proposed dark matter experiments in this mass range~\cite{Higgins:2023gwq,Kalia:2024eml}.
However, as shown below, the modest shielding we assume in our work means that the expected signal in our setup is much larger than was assumed in those works, and we find that the tradeoff between signal and noise that comes with having modest shielding of the apparatus is beneficial to the overall sensitivity (see Fig.~\ref{fig:reach}).

\section{Dark Matter Signals}. 
Various dark matter models can generate small magnetic fields that are potentially detectable with the trapped-ion interferometer technique. The two of particular interest in the mass range to which the technique is most sensitive are kinetically-mixed dark photons and axion-like particles. 

\subsection{Dark Photon Dark Matter.} 
Kinetically-mixed dark photon dark matter is defined by the Lagrangian $\mathcal{L} \supset \tfrac{\epsilon}{2} F_{\mu\nu}F'^{\mu\nu} + \tfrac{1}{2}m_{A'}^2 A'_\mu A'^\mu$, where the first term contains the coupling to the SM photon, and the second endows the dark photon with a mass. The SM and dark photon system can be rotated to the so-called ``interaction'' basis, which has the benefit of making it clear that dark photons couple to the SM through EM fields generated by an effective current, $\boldsymbol{j}'_{\rm eff} = - \epsilon m_{A'}^2 \boldsymbol{A'}$. The dark photon vector potential has a Rayleigh-distributed amplitude, but is on average given approximately by
\begin{align}
\boldsymbol{A'} \sim \mathbb{R}\left[\frac{\sqrt{2\rho_{_{\rm DM}}}}{\sqrt{3} m_{A'}} e^{i m_{A'}t-i \boldsymbol{k}\cdot \boldsymbol{x}}\sum_i \hat{\boldsymbol{n}}_i\, e^{i \varphi_i} \right ] 
\ ,
\end{align}
where $\hat{\boldsymbol{n}}_i$, $i = 1,\,2,\,3$, is a set of orthonormal basis vectors, and $\varphi_i$ are random phases drawn from a flat distribution in $[0,2\pi)$. The scalar potential $A_0' \sim v |\boldsymbol{A'}|$ is suppressed relative to the vector potential, since the Proca condition fixes $\partial_\mu A'^\mu = 0$, and $|\boldsymbol{k}| \sim m_{A'}v$ with $v \sim 10^{-3}\,c$ for virialised dark matter in the Milky Way.

For the mass range we consider here, the dark matter Compton wavelength $\sim 1/m_{A'}$ is very long compared to all scales in the experiment, such as the size of the detector, the size of the shielded room hosting the detector, and even Earth's radius $R_\oplus$.
Therefore, we can work in the quasi-magnetostatic limit to estimate the size of the DM-generated magnetic field, as terms involving time derivatives such as $\partial_t \boldsymbol{E}$  will be sub-dominant relative to terms with spatial derivatives in Maxwell's equations.
Using Amp\`ere's law, we estimate the DP-induced magnetic field from $\nabla\times \boldsymbol{B}_{_{\rm DP}} \simeq \boldsymbol{j}_{\rm eff}$, where the operator $\nabla$ will set the spatial scale over which the field is present.
We therefore expect a typical DM-induced magnetic field magnitude of $B_{_{\rm DP}} \sim - \epsilon m_{A'}^2 L A'$. As we shall see, the scale $L$ that is appropriate for the trapped-ion interferometer is given by the Earth's radius $R_\oplus$.
If we had not accounted for boundary conditions and solved the full wave equation for $B_{_{\rm DP}}$, we might have estimated the DM-induced magnetic field as $B_{_{\rm DP}} \sim \epsilon\,m_{A'} \boldsymbol{v} \times \boldsymbol{A'}$. This na\"ive result is only valid in the regime when $m_{A'}v R_\oplus \gg 1$, which is not the situation considered in this analysis.
The full expression for the dark photon-induced magnetic field with the Earth as a boundary was computed in Ref.~\cite{Fedderke:2021aqo}, and is reproduced in Appendix~\ref{app:magnetic} for completeness. 
Importantly, careful computation suggests that the dominant magnetic field component is tangential to Earth's surface (i.e., the DP-sourced magnetic field has nonzero components only along the $\hat{\boldsymbol{\phi}}$ and $\hat{\boldsymbol{\theta}}$ directions).

\subsection{Axion Dark Matter.} 
For axion or ALP dark matter with a photon interaction, $\mathcal{L} \supset -\tfrac{1}{4}g_{a\gamma\gamma} a F_{\mu\nu}\tilde{F}^{\mu\nu}$, EM fields are generated by an effective current $\boldsymbol{j}_{\rm eff}^a \simeq - g_{a\gamma\gamma} \partial_t a \boldsymbol{B}_0$, to leading order in the DM velocity $v$. 
The axion field is given approximately by
\begin{align}
    a \sim \mathbb{R}\left[ \frac{\sqrt{2\rho_{_{\rm DM}}}}{m_a} e^{i m_a t - i \boldsymbol{k}\cdot \boldsymbol{x} + i \varphi} \right ]
    \ ,
\end{align}
where $\varphi$ is a random phase.
From the form of the effective current, it is clear that a background magnetic field is required to induce a signal. The ion is trapped in an RF Paul trap characterised by a DC and AC potential, such that there is no significant background magnetic field. 
Therefore, to leading order in the DM velocity, there is no signal arising from the axion interacting with the trap EM fields. Instead, an effect may arise from the interaction of the axion with Earth's magnetic field, which induces a signal magnetic field whose magnitude is approximately $B_{a} \sim (g_{a\gamma\gamma} a) (m_a R_\oplus) B_\oplus$ in the $m_a R_\oplus \ll 1$ limit, where $B_\oplus \sim 0.5
\,\text{G}$ is the typical magnitude of Earth's magnetic field. The full axion-induced magnetic field resulting from coupling to Earth's magnetic field was computed in Ref.~\cite{Arza:2021ekq}, and is reproduced in Appendix~\ref{app:magnetic} for completeness. Crucially, as in the DP case, the magnetic field sourced by ALPs is tangential to the Earth's surface.

\subsection{Boundary Conditions Dictating Dark Matter Signals.} 
The statements above about the magnitude of the expected DM-induced magnetic fields are contingent on applying the appropriate boundary conditions to the problem. 
A possible concern is that having a conductive boundary whose length scale $R$ is smaller than $R_\oplus$ might further reduce the signal fields by a factor $R/R_\oplus \ll 1$. 
However, the signals we are interested in measuring are typically in the frequency range $f \lesssim 10^2\,\text{Hz}$. 
Magnetic fields at such low frequencies are notoriously difficult to shield~\cite{Jackson:1998nia}. 
In a Paul trap, the ion is not fully enclosed by conductive metal. Therefore, there is magnetic field leakage into the region where the ion propagates.
The laboratory hosting the ion trap might have a conductive boundary to reduce high-frequency EM noise. 
Computing the shielding efficiency of the precise geometry of the lab is beyond the scope of this work.
However, intuition for the shielding efficiency can be developed by considering a shield with a greater degree of symmetry, such as a sphere. 
As described in detail in Appendix~\ref{sec:shielding}, the modulus of the shielding efficiency, defined as $\eta \equiv |\boldsymbol{B}_{\rm shield}/\boldsymbol{B}_{\rm naive}|$, depends on the sphere's radius $R$, the thickness of the shield $\Delta$, the relative permeability of the shield $\mu$, and $\gamma \simeq (1+i)/\delta_s$, where $\delta_s = \sqrt{2/\omega \sigma \mu}$ is the skin depth. 
In the thin-shield limit, i.e., $\Delta \ll \delta_s, R$, the modulus of the shielding efficiency is approximately
\begin{align}\label{eq:approx_shield_eff}
    \eta \simeq \Bigg\vert 1 + \frac{1}{3\mu} \left (2(\mu-1)^2+R^2 \gamma^2 \right ) \frac{\Delta}{R} \Bigg\vert^{-1} \ .
\end{align}
For typical materials such as copper ($\mu = 1,~\sigma = 2.5\times 10^7\,\text{S/m}$) or ferromagnetic stainless steel ($\mu \sim 1000 ,~\sigma \sim 2\times 10^4\,\text{S/m}$),  the shielding efficiency for a 1 mm thick layer of material in a 5 m radius ``room'' can be approximated using Eq.~\eqref{eq:approx_shield_eff} for $\omega \lesssim 10^{-14}\,\text{eV}$ and is $\eta \gtrsim 0.9$. 
Therefore, unless the thickness of the shielding material is significant, for most of the parameter space we are studying, the shielding does not significantly reduce the signal magnetic field.

However, Earth acts as a good conductor with a significant thickness, as does the ionosphere and interplanetary medium at heights greater than $\sim 300\,\text{km}$ from Earth's surface. Therefore, we may compute the resulting signal fields using the treatment of Refs.~\cite{Fedderke:2021aqo,Arza:2021ekq}, where it was found that the relevant scale is given by $R_\oplus$ as claimed above.

\section{Noise Sources}.
A corollary of the poor shielding of low-frequency signals is that we anticipate ambient magnetic field noise to be the dominant limitation on the ion interferometer sensitivity for DM masses $m_{_{\rm DM}} \lesssim 10^{-14}\,\text{eV}$. 
The expected noise level at $f=1$~Hz is $B_n \simeq 10^{-11\pm1}\,\text{T/}\sqrt{\text{Hz}}$ and scales as $1/f$ across the entire frequency range of interest~\cite{CONSTABLE2023107090,Fullekrug}.~\footnote{Although measured noise levels in certain labs can be up to a factor of 10 higher at similar frequencies, see, e.g.,~\cite{2025arXiv250321725C}.}
These noise scalings appear in our sensitivity projections, with an order of magnitude error applied to account for temporal variations.
Part of this ambient magnetic field noise could be due to vibrations of the apparatus, e.g., time-dependent changes in the orientation of $\hat{\boldsymbol{\Sigma}}$ relative to slow-varying ambient magnetic fields, which mimic a signal at higher frequencies.
%
To mitigate this, one could conceivably run two versions of the experiment alongside one another, with only one being seismically isolated.
A noise subtraction could then be performed by comparing the two datasets.

In addition to ambient noise, we anticipate that shot noise should be the ultimate noise floor for masses above $m_{_{\rm DM}} \gtrsim 10^{-14}\,\text{eV}$. 
For a single ion in the trap, we expect the associated phase uncertainty to be $\delta \phi_{\rm shot} = 1/\sqrt{\text{Hz}}$.
The shot-noise phase uncertainty could be reduced by using a Greenberger-Horne-Zeilinger (GHZ) state~\cite{Greenberger:1989tfe,Greenberger:1990uox} of $N_\mathrm{GHZ}$ ions, in which case the uncertainty would be reduced by $1/N_{\rm GHZ}$.
This phase uncertainty can be mapped onto an equivalent magnetic field noise by using Eq.~\eqref{eq:phase_shift2}. 
This mapping is useful in order to compare signal to noise as real and equivalent magnetic fields.
The equivalent magnetic field of shot noise is $B_{\rm shot} \simeq 5\times 10^{-12} \,  \text{T/}\sqrt{\text{Hz}} \times \mathcal{T}(\omega)^{-1/2} (1\,\mathrm{s}/\Delta t) (1/N_\mathrm{GHZ})$, where $\mathcal{T}(\omega) = \mathrm{sinc}^2(\omega \Delta t/2)$ is the detector's response function.
Here, we have used typical trap parameters so that $|\mathrm{d}\boldsymbol{\Sigma}/\mathrm{d}t| \simeq 5\times 10^{-5}\,\text{m}^2/\text{s}$~\cite{Campbell_2017}.

In order to achieve shot noise-limited sensitivity, various systematic effects associated to the trap itself must be controlled or mitigated.
These have been discussed in detail in Ref.~\cite{West:2019xio}, so we only briefly summarise them here.
Systematic effects typically affect either the phase, the enclosed area or the visibility.
Such effects include differences in the spin-dependent kicks applied to the two ion states, spurious motion of the ion states in the trap, anharmonicities of the trap, and cross-coupling of the trap principal axes.
Of these, it is control of the trap potential that is most crucial for ensuring the performance of the ion interferometer. 
For example, trap anharmonicities of the form $V(x,y) \supset \kappa_x x^2 + \kappa_y y^2+ \lambda_x x^4 + \lambda_y y^4$ should be minimised. 
If $\lambda_i/\kappa_i \sim 0.1$, this would lead to a $10\,\%$ loss in visibility and a modification of the subtended area by the same amount.
As the expected ratios $\lambda_i/\kappa_i \sim 10^{-4}$~\cite{West:2019xio}, only minimal losses in visibility and changes in the interferometer area are expected.
We provide a more detailed discussion of systematic effects in Appendix~\ref{app:imperfections}.
In our sensitivity estimates, we assume they can be controlled at the level required to achieve shot noise-limited sensitivity.

\begin{figure*}[t!]
    \centering
    \includegraphics[width=0.48\linewidth]{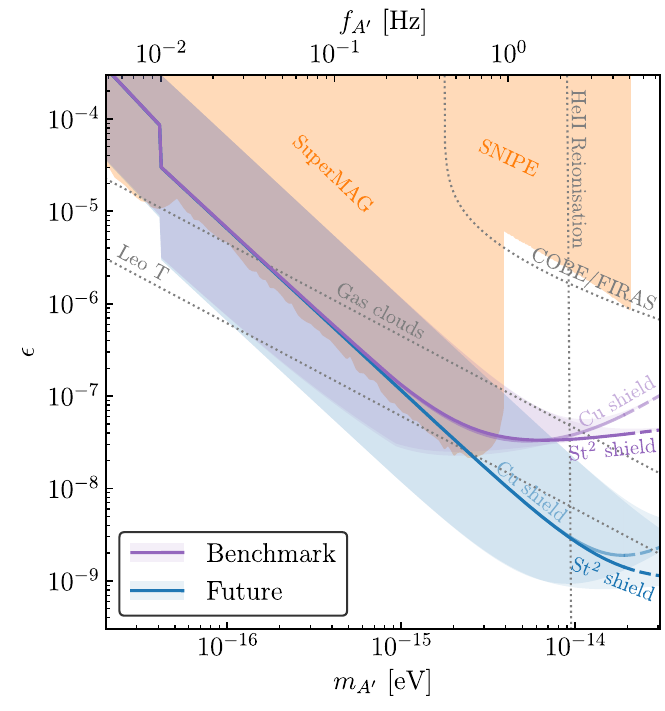}
    \includegraphics[width=0.49\linewidth]{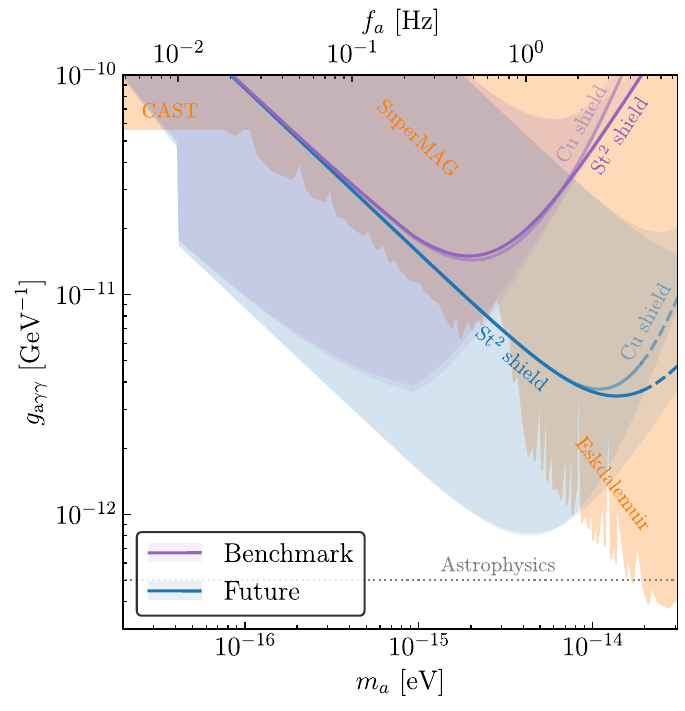}
    \caption{Projected 95\% upper limits on the DP kinetic mixing parameter $\epsilon$ (left panel) and the ALP-photon coupling $g_{a\gamma\gamma}$ (right panel) for a trapped-ion interferometer assuming $T_\mathrm{int} = 10^8$~s and $\rho_{_\mathrm{DM}} = 0.3$~GeV/cm\textsuperscript{3}. The benchmark and optimistic projected sensitivities are shown by purple and blue curves, respectively. The purple and blue shaded regions bound the estimated ambient magnetic noise. At frequencies above 1~Hz, the signal is suppressed by the possible presence of copper (Cu) or stainless steel (St\textsuperscript{2}) shielding around the experiment. The dashed solid lines mark regions of parameter space where the quasistatic approximation is invalid. Regions of parameter space excluded by existing laboratory probes are shaded in orange. Astrophysical and cosmological constraints are shown with dotted lines.}
    \label{fig:reach}
\end{figure*}

\section{Projected sensitivity}. 
In Figure~\ref{fig:reach} we plot the projected power-averaged 95\% upper limits on $\epsilon$ (left panel) and $g_{a\gamma\gamma}$ (right panel) for $T_\mathrm{int}=10^8$~s. As explained in detail in Appendix~\ref{supp:signal}, the 95\% upper limits on the couplings correspond to $\mathrm{SNR} \approx 1.6$ for $T_\mathrm{int}\gg \tau_c$ and $\mathrm{SNR} \approx 12.5$ for $T_\mathrm{int}\ll \tau_c$, where $\tau_c = 2\pi/mv^2$ is the coherence time of the DM field. In purple, we show the projected benchmark sensitivity of a single-ion trap employing the parameters proposed in Ref.~\cite{Campbell_2017}. In particular, we envisage employing $^{171}$Yb$^+$ and the (magnetically insensitive) internal ground hyperfine states $\ket{\downarrow}=\ket{F=0,m_F=0}$ and $\ket{\uparrow}=\ket{F=1,m_F=0}$ with frequency splitting $\omega_a/2\pi \approx 12.6$~GHz. The coherence time of the $\ket{\uparrow}$ state exceeds $10^3$~s~\cite{Fisk}, which allows for an interrogation time $\Delta t = 1$~s. To drive transitions between qubit states, we assume the demonstrated two-photon stimulated Raman transitions using a laser at $355$~nm, which implies a momentum transfer $k_{\rm eff} = 4\pi/355 ~\mathrm{nm}^{-1}$ to a wavepacket's centre-of-mass per SDK~\cite{Campbell:2010und}. In light of their high fidelity, we consider $N=100$ SDKs. Furthermore, we take the non-adiabatic displacement of the trap centre to be $y_d = 100~\mu$m, which can be achieved by nano-second switching of the trapping potentials~\cite{Alonso_2016}, and has recently been shown to yield displacements of $10~\mu$m~\cite{Saito:2021xxd}.

In blue, we show the optimistic projected sensitivity of a trapped-ion interferometer operating with 50 ions prepared in a GHZ state, somewhat beyond the current record of 32 ions entangled in a GHZ state~\cite{Moses2023}, and otherwise identical trap parameters. Equivalently, the sensitivity of the optimistic scenario could be achieved by a single-ion experiment with a fifty-fold enhancement in $|\mathrm{d}\boldsymbol{\Sigma}/\mathrm{d}t|$.

In evaluating both the ALP and DP signals, we assume that the trap's orientation is fixed during the totality of the measurement campaign. Since the DP and ALP signals are tangential to the Earth's surface because of boundary conditions, we assume that $\hat{\boldsymbol{\Sigma}}$ has non-zero components along the $\hat{\boldsymbol{\phi}}$ and $\hat{\boldsymbol{\theta}}$ directions. Since the directionality of the ALP signal is set by the direction of the Earth's magnetic field, which is known, we assume that the trap is oriented so as to be maximally sensitive to the UDLM-sourced magnetic field. In the DP case, because the directionality of the signal is set by the field's polarization, which in models of unpolarized DM varies between coherence patches, we assume for simplicity and without loss of generality that $\hat{\boldsymbol{\Sigma}} \propto \hat{\boldsymbol{\phi}}$. Furthermore, for concreteness, we fix the location of the experiment to be in Los Angeles ($34^\circ \, \rm{N}, 118^\circ \, \rm{W}$).

The blue and purple shaded regions bound the magnitude of magnetic ambient noise in optimistic and pessimistic scenarios. To illustrate the impact of laboratory shielding on the signal, we also present the projected sensitivity when the tabletop experiment is enclosed within a hollow copper or stainless steel sphere with a radius of 5~m and a wall thickness of 1~mm in lighter and darker shades, respectively. In the limit $m_{_{\rm DM}} R_\oplus \lesssim 1$, where the spatial and conducting features of the Earth's core affect the signal modeling, we plot the projected sensitivity with dashed lines. 

The regions of parameter space shaded in orange are excluded by existing DM laboratory searches, such as SuperMAG~\cite{Fedderke:2021aqo,Arza:2021ekq,Fedderke:2021rrm,Friel:2024shg}, SNIPE Hunt~\cite{Sulai:2023zqw}, AMAILS~\cite{Jiang:2023jhl}, Eskdalemuir~\cite{Nishizawa:2025xka,Taruya:2025zql} and CAST~\cite{CAST:2024eil}. We show constraints from astrophysical or cosmological probes with dotted lines. These include anomalous heating of the Leo T dwarf galaxy, the Milky Way gas cloud G33.4-8.0~\cite{Wadekar:2019mpc} and the intergalactic medium (IGM) around the epoch of HeII reionization~\cite{Caputo:2020bdy}, as well as spectral distortions of the CMB as measured by FIRAS aboard COBE~\cite{McDermott:2019lch,Caputo:2020bdy,Arsenadze:2024ywr,Chluba:2024wui} and of the X-ray spectrum of the quasar H1821+643 as measured by Chandra~\cite{Reynes:2021bpe}.

\section{Discussion and conclusions}.
%
In this work we have studied the prospects of using trapped ions as matter-wave interferometers to search for electromagnetically-coupled ultralight dark matter,
which imprints a non-zero Aharonov-Bohm-like phase on spin states of the ion.

The proposed protocol involves applying $N$ spin-dependent kicks in momentum space to an ion in a Schr\"odinger cat state: a superposition of spatially-delocalised up and down spin states, effectively entangling spin and position degrees of freedom.
This protocol has the benefit of enhancing the sensitivity to magnetic fields by causing the ion spin states to subtend a significantly larger effective area than if they remained in their ground states of motion.
Enormous improvements in the manipulation and control of trapped ions have mostly been driven by potential applications in quantum computing.
Here, we demonstrate that these also have interesting applications for quantum sensing of dark matter.

Taking parameters for such a trapped-ion interferometer as proposed originally in Ref.~\cite{Campbell_2017}, we have demonstrated that a single ion could be used to probe new parameter space for dark photon and axion-like particle dark matter.
The peak sensitivity is obtained for dark matter mass $m_{_{\rm DM}} \sim 1/\Delta t$, where $\Delta t$ is the interrogation time of the ion, which can be as long as a second.
The sensitivity at smaller masses is limited by the fact that electromagnetic shielding is not expected to be sufficient to prevent ambient magnetic field noise from penetrating into the region containing the ion trap.
At masses larger than $m_{_{\rm DM}} \sim 1/\Delta t$, the sensitivity decreases owing to the fact that many complete oscillations of the dark matter field are integrated over during the ion interrogation time, leading to a partial cancellation of the signal phase.

This proposal is complementary to existing searches for dark photon and axion-like particle dark matter such as SuperMAG~\cite{Fedderke:2021aqo,Fedderke:2021rrm,Arza:2021ekq,Friel:2024shg}, SNIPE~\cite{Sulai:2023zqw} and AMAILS~\cite{Jiang:2023jhl} that utilize networks of magnetometers all over the world, and an analysis~\cite{Nishizawa:2025xka,Taruya:2025zql} of data from the Eskdalemuir magnetometry station in the UK.

As seen in Fig.~\ref{fig:reach}, the trapped-ion interferometer could be an important laboratory probe of regions of the parameter spaces for both dark matter candidates,
beyond those constrained by astrophysical or cosmological considerations~\cite{Wadekar:2019mpc,McDermott:2019lch,Caputo:2020bdy,Arsenadze:2024ywr,Chluba:2024wui,Marsh:2017yvc,Reynolds:2019uqt,Dessert:2020lil,Reynes:2021bpe,Hoof:2022xbe,Ning:2024eky}.
Additionally, preparing multiple ions in an entangled GHZ state~\cite{Greenberger:1989tfe,Greenberger:1990uox} would lead to a linear improvement in the sensitivity to the dark matter coupling, as shown in Fig.~\ref{fig:reach}. 
While we have considered 50 entangled ions explicitly in our sensitivity estimate, the needs of quantum computers are such that preparing multi-ion GHZ states is an area of active research~\cite{PhysRevLett.106.130506,Moses2023}.
Bearing in mind the possibility of protocols for entangling thousands of states~\cite{Zhao2021}, the potential for improving the sensitivity of our detector concept is substantial.

\begin{acknowledgments}

We thank Wesley Campbell, 
Jack Devlin, 
Paul Hamilton,
Saarik Kalia and Nicholas Rodd for their helpful comments.
The work of LB was supported by the U.S. Department of Energy, Office of Science, Office of High Energy Physics under Award Number DE-SC0011632, and by the Walter Burke Institute for Theoretical Physics. 
DB acknowledges support from the Departament de Recerca i Universitats from Generalitat de Catalunya to the Grup de Recerca 00649 (Codi: 2021 SGR 00649).
This publication is part of the R\&D\&i project PID2023-146686NB-C31 funded by MICIU/AEI/10.13039/501100011033/ and by ERDF/EU.
IFAE is partially funded by the CERCA program of the Generalitat de Catalunya.
This work is supported by ERC grant ERC-2024-SYG 101167211. Funded by the European Union. Views and opinions expressed are however those of the author(s) only and do not necessarily reflect those of the European Union or the European Research Council Executive Agency. Neither the European Union nor the granting authority can be held responsible for them.
The work of JE was supported by the United Kingdom STFC Grants ST/X000753/1 and ST/T000759/1.
The work of SARE was supported by SNF Ambizione grant PZ00P2\_193322, \textit{New frontiers from sub-eV to super-TeV}.
DB and SARE thank the Mainz Institute for Theoretical Physics, where this work was completed, for their hospitality.

\end{acknowledgments}

\appendix

\section{Trapped-Ion Interferometry Protocol}\label{app:protocol}

In this appendix, we review the trapped-ion interferometry protocol suggested in Ref.~\cite{Campbell_2017}, and calculate the signal induced by the dark matter magnetic fields.

\subsection{State preparation and displacement operators}

The initial ion state is a coherent state of motion and a pure spin state, i.e.
\begin{align}
    \ket{\psi_i} = \ket{\alpha_x,\alpha_y}\otimes \ket{\uparrow} \equiv \ket{\alpha_x,\,\alpha_y;\uparrow} \ .
\end{align}
The coherent position state means that $\bra{\psi_i}\hat{x}\ket{\psi_i} = 2x_0 \mathbb{R}[\alpha_x]$, and $\bra{\psi_i}\hat{p}_x\ket{\psi_i} = 2 m \omega_xx_0 \mathbb{I}[\alpha_x]$, where $x_0 = \sqrt{\tfrac{\hslash}{2 M \omega_x}}$.

\begin{figure*}
    \centering
    \includegraphics[width=\textwidth]{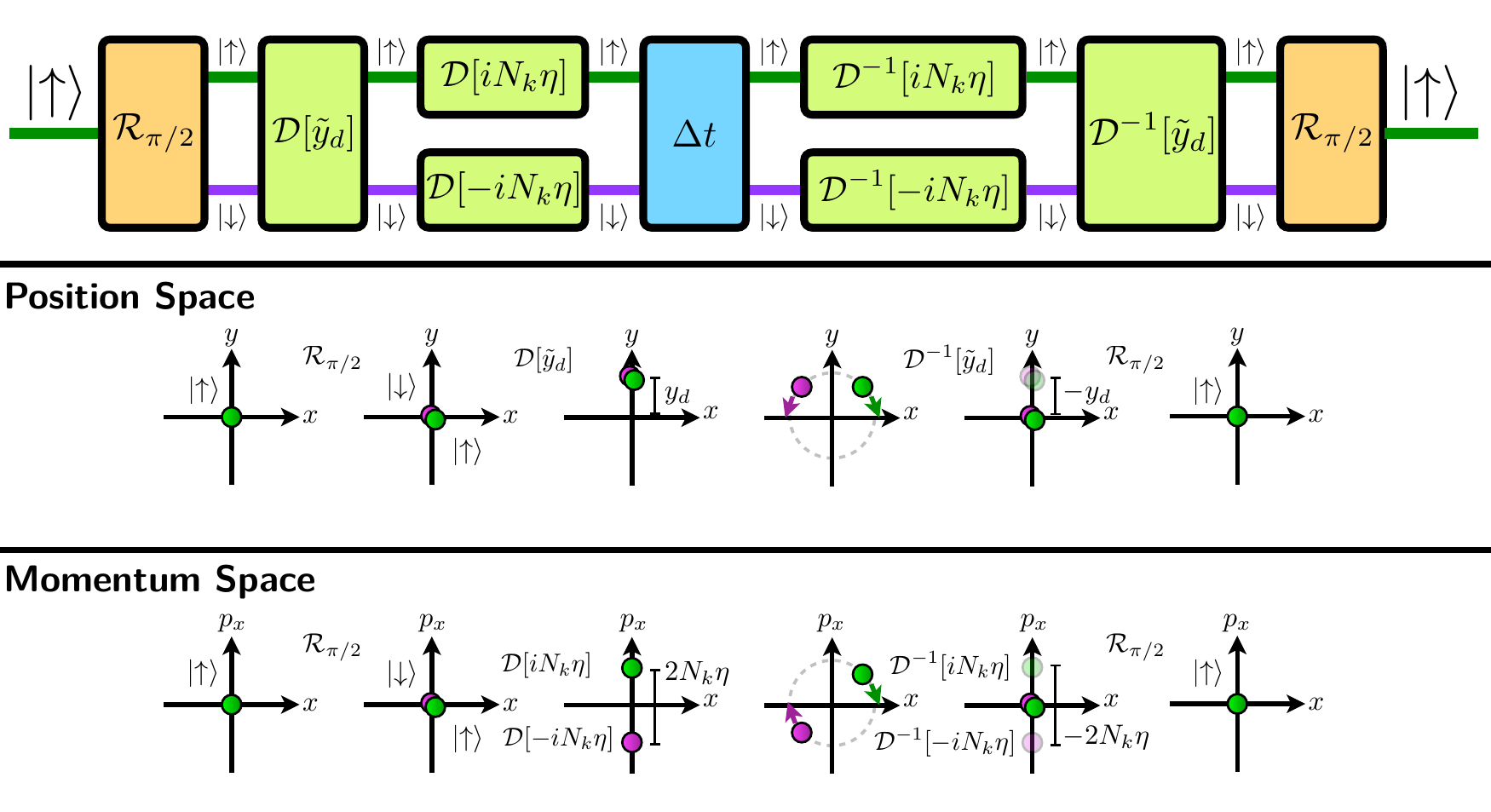}
    \caption{A depiction of the trapped-ion interferometer protocol in terms of operations applied to the states (top row), the resulting motion of the state in position space (middle row), and in $x$ momentum space (bottom row). 
    }
    \label{fig:protocol}
\end{figure*}

The trapped-ion interferometer protocol, depicted in Fig.~\ref{fig:protocol}, consists of applying a $\pi/2$ pulse to separate $\ket{\psi_i}$ into a superposition of up and down spin states, before applying a coherent displacement by an amount $y_d$ in the $y$-direction, followed by spin-dependent kicks in the $p_x$ direction. The superposition of ion states is then allowed to evolve freely for a time $\Delta t$, after which all applied transformations are reversed. Thus, the final state is given by
\begin{equation}
\begin{aligned}
    \ket{\psi_f} & = \underbrace{\mathcal{R}_{\pi/2}\mathcal{D}_y[-\tilde{y}_d]\mathcal{D}_{\rm SDK}^{-1}}_{\text{applied after $\Delta t$}} \\ & \qquad \qquad \times U(\Delta t,0) \mathcal{D}_{\rm SDK} \mathcal{D}_y[\tilde{y}_d] \mathcal{R}_{\pi/2} \ket{\psi_i} \, ,
\end{aligned}
\end{equation}
where $U(\Delta t, 0)$ accounts for the unitary evolution of the state from initialization to time $\Delta t$. The dimensionless displacement is defined here as $\tilde{y}_d = y_d/y_0$.
If no external operation acts on the ion states, then $\ket{\psi_f} = \ket{\psi_i}$ up to a global phase. 
We employ the following notation for rotations,
\begin{align}
    &\mathcal{R}_{\pi/2} \equiv \frac{1}{\sqrt{2}}\bigg[ \Big(\ket{\uparrow} + \ket{\downarrow}\Big) \bra{\uparrow } + \Big(\ket{\uparrow} - \ket{\downarrow}\Big) \bra{\downarrow} \bigg] \ ,
\end{align}
and displacements,
\begin{align}
    & \mathcal{D}[O] \equiv e^{O\hat{a}^\dagger - O^*\hat{a}} = e^{-|O|^2/2}e^{O\hat{a}^\dagger}e^{-O^\star\hat{a}}\ , \\
    & \mathcal{D}_{\rm SDK} = \mathcal{D}[i N \eta] \ket{\uparrow}\bra{\uparrow} + \mathcal{D}[-i N \eta] \ket{\downarrow}\bra{\downarrow} \ .
\end{align}
The second line defines the spin-dependent kick displacement operator in terms of the Dicke parameter $\eta \equiv k_{\rm eff} x_0$. The quantity $k_{\rm eff}$ is the momentum transferred to the spin state in each kick.  

\subsection{Aharonov-Bohm Phase as a Displacement Operator}

In the presence of a vector potential coupling to the ion, we can write the Hamiltonian in the Schr\"odinger picture for the system as
\begin{equation}
\begin{aligned}
    \hat{H}_S &= \frac{1}{2 M}\left[\left(\hat{p}_{x,S} - q A_x(t) \right)^2+\left(\hat{p}_{y,S} - q A_y(t) \right)^2\right] \\
    & \qquad \qquad + \frac{1}{2}M \left( \omega_x^2 \hat{x}_S^2 + \omega_y^2 \hat{y}_S^2 \right) \ ,
\end{aligned}
\end{equation}
where 
\begin{align}
\hat{x}_S &= x_0 \left(\hat{a}_{x,S}^\dagger + \hat{a}_{x,S} \right) \, , \\
\hat{p}_{x,S} &= M \omega x_0 \left(\hat{a}_{x,S}^\dagger - \hat{a}_{x,S} \right) \, , 
\end{align}
and similarly for $y$.
With these definitions, we can easily write the Hamiltonian as the usual free term 
\begin{equation}
\hat{H}_{0,S} = \hslash \omega_x \hat{a}_{x,S}^\dagger \hat{a}_{x,S}+ \hslash \omega_y \hat{a}_{y,S}^\dagger \hat{a}_{y,S}
\end{equation}
and an interaction Hamiltonian
\begin{equation}
\begin{aligned}
    \hat{H}_{\text{int},S} &= i q \Big [ x_0 \omega_x\left( A_x^*(t) \hat{a}_{x,S} - A_x(t) \hat{a}_{x,S}^\dagger \right) \\ 
    & \qquad + y_0 \omega_y \left( A_y^*(t) \hat{a}_{y,S} - A_y(t) \hat{a}_{y,S}^\dagger\right) \Big ] \ .
\end{aligned}
\end{equation}
In the interaction picture, this Hamiltonian corresponds to a unitary evolution operator of the form 
\begin{equation}
    U_{x,I}(t,0) = \text{exp}[\beta_x(t) \hat{a}_{x,S}^\dagger - \beta_x^*(t) \hat{a}_{x,S} ],
\end{equation} 
which acts on the position state $x$. Once again, a similar expression can be written for $y$. In the above expression,
\begin{align}
    \beta_x(t) = - \frac{x_0 \omega_x q}{\hslash} \int_0^t \mathrm{d}t'\,A(t) e^{i \omega_x t'} \ .
\end{align}
We could then perform the rest of the calculation in terms of components of the vector potential. However, we can simplify the calculation by writing the vector potential in terms of the magnetic field in symmetric gauge.
For an external magnetic field $\boldsymbol{B}_{\rm ext} = B_{\rm ext}\hat{z}$, we find that we can write the evolution operator as
\begin{align}
    &\mathcal{D}_{\rm AB} = \text{exp} \left[\frac{i}{\hslash} \beta_z(t) \hat{J}_z \right] \ , 
\end{align}
where
\begin{align}
    & \hat{J}_z = i \hslash (\hat{a}_{x,S} \hat{a}_{y,S}^\dagger - \hat{a}_{y,S}\hat{a}^\dagger_{x,S}) \ , \\ 
    & \beta_z(t) = \frac{q}{2M}\int_0^t \mathrm{d}t'B_{\rm ext}(t') \ .
    \label{supp:eq:beta_z}
\end{align}
To obtain the displacement operator $\mathcal{D}_{\rm AB}$ in this form, we have assumed that $\omega_x = \omega_y$, so that $x_0 = y_0$ also. We see that this means the effect of the magnetic field is to perform a rotation about $\hat{J}_z$ of the ion states.

To compute the effect of the external magnetic field on the ion states during the trapped-ion interferometry sequence, we can transform the displacement operators applied to the ion states as $\mathcal{D} \to \mathcal{D}' = \mathcal{D}^{-1}_{\rm AB} \mathcal{D} \mathcal{D}_{\rm AB}$.
For the coherent displacement operator along $y$ and the spin-dependent kick operators, we find 
\begin{align}
    \mathcal{D}'[-\tilde{y}_d] &= \text{exp}\Big[- \tilde{y}_d \left(\cos\beta_z \hat{a}_y^\dagger - \sin\beta_z \hat{a}_x^\dagger \right) \nonumber \\ &  \qquad \qquad  + \tilde{y}_d \left(\cos \beta_z \hat{a}_y - \sin \beta_z \hat{a}_x \right)\Big] \ , \\
    \mathcal{D}'[i N \eta] &= \text{exp}\Big[i N \eta \left(\cos\beta_z \hat{a}_x^\dagger + \sin\beta_z \hat{a}_y^\dagger \right) \nonumber \\
    & \qquad \qquad - i N \eta \left(\cos\beta_z \hat{a}_x + \sin\beta_z \hat{a}_y \right) \Big] \ .
\end{align}
As expected, the Aharonov-Bohm phase leads to mixing of directions when a displacement operator is applied.
Since $\hat{a}_x$ and $\hat{a}_y$ commute, the displacement operators can be divided into parts applied along the $x$ and $y$ directions separately.
This leads us to define the quantity $\tilde{y}_d^x = y_d/x_0$ as the $y$-direction trap displacement normalised to the zero-point motion in the $x$-direction.
This quantity is always accompanied by a $\beta_z$-dependent function. 

\subsection{Aharanov-Bohm Effect on Trapped-Ion Interferometry}

With the definitions introduced above, we can now derive the effect of the external magnetic field on the trapped ion through the application of successive displacement operators as
\begin{align}
    \ket{\psi_f} = \mathcal{R}_{\pi/2}{\mathcal{D}_y'}[-\tilde{y}_d]{\mathcal{D}'}_{\rm SDK}^{-1} \mathcal{D}_{\rm SDK} \mathcal{D}_y[\tilde{y}_d] \mathcal{R}_{\pi/2} \ket{\psi_i} \ ,
\end{align}
where we have assumed that the duration of the first displacements on the state is sufficiently short that they can be treated as being unperturbed by $B_{\rm ext}$. Meanwhile, the long duration between the application of $\mathcal{D}_{\rm SDK}$ and $\mathcal{D}^{-1}_{\rm SDK}$ means that only the latter and subsequent displacements are perturbed.

Prior to the application of the final $\pi/2$ pulse, the ion state can be described by $\ket{\psi} = \ket{\psi_{\uparrow}} \otimes \ket{\uparrow} + \ket{\psi_{\downarrow}} \otimes \ket{\downarrow} $,
where the definitions of $\ket{\psi_{\uparrow, \downarrow}}$ facilitate our subsequent analysis.
In the presence of the signal associated to $\mathcal{D}_{\rm AB}$, we expect
\begin{align}
    \ket{\psi_{\uparrow}} &\equiv e^{i \delta} \mathcal{D}_x [\alpha_x + i N \eta(1-c_{\beta_z}) + \tilde{y}_d^x s_{\beta_z}] \ket{0}_x \nonumber \\
    & \quad \otimes \mathcal{D}_y [\beta_y + \tilde{y}_d(1-c_{\beta_z}) - i N \eta s_{\beta_z}] \ket{0}_y \ , \\
    \ket{\psi_{\downarrow}} &\equiv e^{-i \delta} \mathcal{D}_x [\beta_x - i N \eta(1-c_{\beta_z}) + \tilde{y}_d^x s_{\beta_z}] \ket{0}_x \nonumber \\
    & \quad \otimes \mathcal{D}_y [\beta_y + \tilde{y}_d(1-c_{\beta_z}) + i N \eta s_{\beta_z}] \ket{0}_y \ ,
\end{align}
where the displacement operators indicate how to construct the spatial state from the vacuum, and the overall phase $\delta$ is given by
\begin{equation}
\begin{aligned}
    \delta & = N \eta \mathbb{R}[\alpha_x](1-c_{\beta_z}) \\
    & \quad - N \eta(\tilde{y}^x_d +\tilde{y}_d - (\tilde{y}_d-\tilde{y}^x_d)c_{\beta_z} + \mathbb{R}[\alpha_y])s_{\beta_z}\ .
    \label{supp:eq:nontrivialphase}
\end{aligned}
\end{equation}
After the final $\mathcal{R}_{\pi/2}$, we can therefore write $\ket{\psi_f} = \tfrac{1}{2} \left(\left( \ket{\psi_\uparrow} + \ket{\psi_\downarrow}\right) \otimes \ket{\uparrow} + \left(\ket{\psi_\uparrow} - \ket{\psi_\downarrow}\right) \otimes \ket{\downarrow} \right)$, with which we can evaluate the probability of measuring a given spin state as
\begin{align}
    \mathcal{P}[\uparrow] = \int \mathrm{d}^2\alpha_x \mathrm{d}^2 \alpha_y P(\alpha_x) P(\alpha_y) \big\vert \braket{\uparrow|\psi_f}\big\vert^2 \ .
\end{align}
The functions $P(\alpha_{x,y})$ are the Glauber-Sudarshan $P$-representations of the initial motional state in the coherent state basis \cite{Mandel:1995seg}.
An analogous expression can be written for the probability of measuring a $\ket{\downarrow}$ state.
For a symmetric trap, i.e., one where $\omega_x = \omega_y$ and $x_0 = y_0$, we can therefore simplify the quantity $|\braket{\uparrow | \psi_f}|^2$ as
\begin{widetext}
\begin{align}
    |\braket{\uparrow | \psi_f}|^2 &=\frac{1}{2} \Bigg(1 + e^{-2(2N \eta \sin(\beta_z/2))^2} \cos\left( 4N \eta  \mathbb{R}[\alpha_x](1-\cos{\beta_z}) - \frac{A(\alpha_y)}{\pi x_0y_0}\sin{\beta_z}\right) \Bigg) \ ,
    \label{eq:finalproduct}
\end{align}
\end{widetext}
where we have also defined $A(\alpha_y) = 4\pi x_0 N \eta \left(y_0 \mathbb{R}[\alpha_y] + y_d \right)$ as the effective area subtended by the ion motion. To leading order in small $\beta_z$, we  see that the argument of the cosine is enhanced by the ratio $A(\alpha_y)/\pi x_0 y_0 \simeq 4 N \eta \tilde{y}_d \gg 1$, reflecting the enhanced phase sensitivity of the ion-interferometer protocol.

The result of Eq.~\eqref{eq:finalproduct} is precisely of the usual form expected in matter-interferometric measurements, namely $|\braket{\uparrow |\psi_f}|^2 = \frac{1}{2}(1+\mathcal{V}\cos\varphi)$, where $\mathcal{V}$ is the visibility and $\varphi$ is the phase. In the small-$\beta_z$ limit we have
\begin{align}
    \mathcal{V} &= 1 +\mathcal{O}(\beta_z^2) \ , \\
    \mathcal{\varphi} &= -4 N \eta \tilde{y}_d \beta_z + \mathcal{O}(\tilde{y}_d^0\,\beta_z^2) \ .
    \label{supp:eq:signalphase}
\end{align}
In the second line, we have only retained the leading order term in $\beta_z$ and in $\tilde{y}_d \gg 1$.

We wrote in the main text in Eq.~\eqref{eq:phase_shift} that the expected phase shift depends on the time integral of $\boldsymbol{B}\cdot \mathrm{d}\boldsymbol{\Sigma}/\mathrm{d}t$.
Mapping that result to Eq.~\eqref{supp:eq:signalphase}, we see that we can identify the relation between $\mathrm{d}\boldsymbol{\Sigma}/\mathrm{d}t$ and the trap parameters such as $N,~\eta,~\tilde{y}_d$ through the definition of $\beta_z$ in Eq.~\eqref{supp:eq:beta_z}.
Differentiating both Eq.~\eqref{eq:phase_shift} and Eq.~\eqref{supp:eq:signalphase} with respect to time, we can see that we can identify
\begin{align}
    \Bigg\vert\frac{\mathrm{d}\boldsymbol{\Sigma}}{\mathrm{d}t}\Bigg\vert = \frac{N \eta y_d}{y_0 m_{\rm ion}} =\frac{N k_{\rm eff} y_d}{m_{\rm ion}} \ , 
\end{align}
assuming $x_0 = y_0$, as claimed in the main text.

\subsection{Effect of Imperfections}\label{app:imperfections}

Here we discuss the effects of systematic errors in the trap parameters and ion manipulation operations on the sensitivity.
These systematic errors can be evaluated through their effect on the visibility $\mathcal{V}$ and the observable phase $\varphi$.
Another possible source of difficulties is thermal phonons.
We discuss this possibility first, and then discuss trap systematic effects.

\subsubsection{Thermal Phonons}

If there is a non-zero thermal occupation of phonon modes $\bar{n}$ in the trap, the Glauber-Sudarshan $P$-representation that must be integrated over is
\begin{align}
    P(\alpha_i) = \frac{e^{-|\alpha_i|^2/\bar{n}_i}}{\pi \bar{n}_i} \ ,
\end{align}
where the occupation numbers $\bar{n}_i$ in the $x$ and $y$ directions may differ.
When evaluating the probability $\mathcal{P}[\uparrow]$ with these $P$-representations, we find
\begin{widetext}
\begin{align}
    \mathcal{P}[\uparrow] = \frac{1}{2}\left\{1 + \exp\bigg[-2(2N\eta\sin(\beta_z/2))^2\left(1+\bar{n}_x+\bar{n}_y + (\bar{n}_y-\bar{n}_x) \cos\beta_z\right)\bigg]
    \,\cos\left(\frac{4 N\eta\, y_d\, \sin\beta_z}{x_0}\right)
    \right \}  \ .
\end{align}
\end{widetext}
We see that the visibility reduces to ${\mathcal{V} = e^{-2(2N\eta\sin(\beta_z/2))^2\left(1+2\bar{n}\right)}}$ if $\bar{n}_x = \bar{n}_y$ as assumed in Ref.~\cite{Campbell_2017}, which matches their result.

Evidently, the thermal occupation number only affects the visibility, and is only relevant if the combination $\bar{n} \sin(\beta_z/2)^2 \sim \mathcal{O}(1)$. 
For typical expected values of $\beta_z$, the effect is tiny for anticipated values of $\bar{n} \sim \mathcal{O}(100)$.
If the thermal occupation numbers $\bar{n}_x \neq \bar{n}_y$, there is still only a limited effect. 
To leading order in $\beta_z^2$, only $\bar{n}_y$ appears, and $\bar{n}_x$ only enters at $\mathcal{O}(\beta_z^4)$ in the exponent of $\mathcal{V}$.

\subsubsection{Semi-Classical Treatment of Imperfections}

As discussed in the main text, a serious possible source of error is anharmonicity in the trap potential.
In this case, the trap potential is modelled as
\begin{equation}
\begin{aligned}
    V(x,y) & = \kappa_x x^2 + \kappa_y y^2 + \kappa_{xy} x y + \gamma_x x^3 + \gamma_y y^3 \\ & \qquad  + \gamma_{xy} x^2 y + \gamma_{yx} x y^2 + \lambda_x x^4 + \lambda_y y^4 \\ & \qquad + \ldots \ ,
\end{aligned}
\end{equation}
where the $\kappa_{ij},~\gamma_{i(j)},~\lambda_i$ coefficients are treated phenomenologically. 
Typical traps (e.g., Paul traps) have significant symmetry, so that the $\gamma_{i(j)}$ terms can be ignored. 

The remaining imperfections can be treated semi-classically, solving the equation of motion for $x(t),~y(t)$ in the presence of a given potential term.
The effect on the phase can be estimated by evaluating the path integral, i.e., by computing
\begin{align}
    \delta\varphi(t) = \frac{1}{\hslash} \int_0^t \mathrm{d}t' \mathcal{L}(t') \ ,
\end{align}
where $\mathcal{L}(t')$ is the classical Lagrangian of the system. 

We can easily evaluate the phase accumulated by the ion after a time $2 n\pi/\omega_x$ in the trap.
In the unperturbed trap with $\omega_x = \omega_y$ there would be no non-zero phase in the absence of a signal.
If $\omega_x \neq \omega_y$, there would still be no non-zero phase in the absence of a signal due to symmetry.
However, there would be a change in the effective area subtended by the ions, as can be seen from Eq.~\eqref{supp:eq:nontrivialphase}, which can be mapped onto the area $A(\alpha_y)$.
Therefore, in the presence of a signal the area would change, as would the resulting phase, becoming $\varphi_{\omega_x \neq \omega_y} \simeq - 4 N \eta \tilde{y}_d^x \beta_z$, where we recall that $\tilde{y}_d^x = y_d/x_0$.
This differs from the previous result of Eq.~\eqref{supp:eq:phase_shift} by a factor $\sqrt{\omega_x/\omega_y}$.

Let us now consider non-zero $\kappa_{x,y}$ terms. If they are identical, $\kappa_x=\kappa_y$, then the trap remains axisymmetric.
We can identify the effective trap frequencies as $\omega_i^2 \to \omega_i^2 + 2\kappa_i/m$.
This leads to our first observation, namely that we must satisfy $\kappa_i \geq -m \omega_i^2/2$, or else the solution to the ion's equation of motion is divergent at large times. 
With the appropriate redefinition of $\omega_i^2$, and assuming the $\kappa_i$ satisfy the requirement above, we simply find the same result we had for the unperturbed trap described above, albeit with the appropriate redefinition of all $\omega_i$-dependent quantities.

If we introduce a trap cross-coupling, $\kappa_{xy} \neq 0$, then the classical solution to the equation of motion gives
\begin{align}
    x(t) = x_0 \cosh \left[t\left( -\omega_x^2 -\frac{\kappa_{xy}}{m} \right)^{1/2}\right] \ ,
\end{align}
and a similar equation for $y(t)$.
The solution for $x_i(t)$ is such that we can identify a new oscillation frequency $\omega'_i = \sqrt{\omega_i^2 - \kappa_{xy}/m}$.
Integrating for a time $t' = n\pi/\omega'$ leads to no accumulated phase in the interferometer.

Finally, let us consider non-zero quartic terms, $\lambda_i$. 
We can solve the equation of motion classically for small excursions about $x_0,~y_0$ by approximating $x(t)^3 \sim x(t) x_0^2$ and similarly for $y(t)^3$.
We find
\begin{align}
    x(t) \simeq x_0 \cosh\left[t\left(-\omega_x^2 - \frac{4\lambda_x x_0^2}{m}  \right)^{1/2}\right] \ ,
\end{align}
and a similar expression for $y(t)$.
We see that, once again, we can identify a new oscillation frequency $\omega'_i = \sqrt{\omega_i^2+4\lambda_i x_{i,0}^2/m}$ and there will be no non-zero accumulated phase.

Other effects due to these imperfections, such as changes in the ion subtended area and the visibility, are discussed at some length in Ref.~\cite{West:2019xio}.

\section{Magnetic fields}\label{app:magnetic}

\begin{table*}[t!]
 \centering
    \renewcommand{\arraystretch}{1.8
    }
    \resizebox{\textwidth}{!}{
    \begin{tabular}{c|c|c|c|c}
    \hline
    \hline
    \multicolumn{5}{c}{\large $\boldsymbol{\Phi}_{\ell m}$} \\
    \hline
         \large $~~\ell~~$ & \large $m=0$ & \large $m=1$ & \large $m=2$ & \large $m=3$ \\
         \hline
         \large $0$ & $0$ & -- & -- & -- \\
         \large $1$ & $\begin{aligned}-\sqrt{\frac{3}{4\pi}} \sin \theta \, \hat{\boldsymbol{\phi}}\end{aligned}$ & $\begin{aligned} -& \sqrt{\frac{3}{8\pi}} e^{i\phi} 
         \Big (\cos \theta \, \hat{\boldsymbol{\phi}} -i \hat{\boldsymbol{\theta}} \Big ) \end{aligned} $ & -- & -- \\
         \large $2$ & $\begin{aligned}-\sqrt{\frac{45}{4\pi}} \sin \theta \cos \theta \, \hat{\boldsymbol{\phi}}\end{aligned}$ & $\begin{aligned}& \sqrt{\frac{15}{8\pi}}e^{i\phi} 
         \Big ((1-2 \cos^2 \theta )\, \hat{\boldsymbol{\phi}} + i \cos \theta \, \hat{\boldsymbol{\theta}} \Big ) \end{aligned}$ & $\begin{aligned}\sqrt{\frac{15}{8\pi}}e^{2i\phi} \sin \theta \left ( \cos \theta \, \hat{\boldsymbol{\phi}} - i \, \hat{\boldsymbol{\theta}} \right )\end{aligned}$ & -- \\
         \vspace{.1cm}\large $3$ & $\begin{aligned}-& \sqrt{\frac{63}{16\pi}} 
         \left ( 5 \sin \theta \cos^2 \theta - \sin \theta \right ) \hat{\boldsymbol{\phi}} \end{aligned}$ & $\begin{aligned} & \sqrt{\frac{21}{64\pi}} e^{i\phi} \Big (  (5 \cos^3  \theta -9 \cos \theta ) \hat{\boldsymbol{\phi}} 
         + i (5\cos^2 \theta -1 ) \hat{\boldsymbol{\theta}} \Big ) \end{aligned}$ & $\begin{aligned}\sqrt{\frac{105}{32\pi}} e^{2i\phi} \sin \theta \left ((3\cos^2 \theta + 2) \hat{\boldsymbol{\phi}} -2i \cos \theta \hat{\boldsymbol{\theta}} \right )\end{aligned}$ & $\begin{aligned}\sqrt{\frac{315}{64\pi}} e^{3i\phi} \sin^2 \theta (\cos \theta \hat{\boldsymbol{\phi}}-i\hat{\boldsymbol{\theta}})\end{aligned}$ \\
        \hline
        \hline
    \end{tabular}}
    \caption{Vector spherical harmonics $\boldsymbol{\Phi}_{\ell m}$ determined using the definitions provided in Ref.~\cite{Barrera_1985}. By choosing the $\hat{\boldsymbol{z}}$ direction to be aligned with the Earth's axis of rotation, $\phi$ can be identified with the longitude  and $\pi/2-\theta$ with the latitude.}
    \label{tab:Philm}
\end{table*}

\subsection{Vector spherical harmonics conventions} \label{sec:VSH}

Here, we briefly summarise the relevant vector spherical harmonics conventions used in this paper, which agree with the ones adopted in Refs.~\cite{Fedderke:2021aqo,Arza:2021ekq,Barrera_1985}. The vector spherical harmonics are defined in terms of scalar spherical harmonics $Y_{\ell m}$ as
\begin{align}
\boldsymbol{Y}_{\ell m} & = Y_{\ell m} \hat{\boldsymbol{r}} \, , \\ \boldsymbol{\Psi}_{\ell m} &= r \boldsymbol{\nabla} Y_{\ell m} \, , \\ \boldsymbol{\Phi}_{\ell m} &= \boldsymbol{r}\times \boldsymbol{\nabla} Y_{\ell m} \, ,
\end{align}
where $Y_{\ell m}$ are functions of polar and azimuthal angles, i.e., $\Omega=(\theta, \phi)$, and $ -\ell \leq m \leq \ell$ for $\ell \geq 0$. These satisfy the following properties:
\begin{align}
& \boldsymbol{Y}_{\ell,-m}=(-1)^m \boldsymbol{Y}_{\ell m}^* \, , \\
& \boldsymbol{\Psi}_{\ell,-m}=(-1)^m \boldsymbol{\Psi}_{\ell m}^* \, , \\
& \boldsymbol{\Phi}_{\ell,-m}=(-1)^m \boldsymbol{\Phi}_{\ell m}^* \, ,
\end{align}
and the orthogonality relations
\begin{align}
\boldsymbol{Y}_{\ell m} \cdot \boldsymbol{\Psi}_{\ell m} & =\boldsymbol{Y}_{\ell m} \cdot \boldsymbol{\Phi}_{\ell m}=\boldsymbol{\Psi}_{\ell m} \cdot \boldsymbol{\Phi}_{\ell m}=0 \, .
\end{align}
Additionally, the relationships between the Cartesian unit vectors and the vector spherical harmonics are
\begin{align}
& \hat{\boldsymbol{x}}=-\sqrt{\frac{2 \pi}{3}}\left(\boldsymbol{Y}_{11}-\boldsymbol{Y}_{1,-1}+\boldsymbol{\Psi}_{11}-\boldsymbol{\Psi}_{1,-1}\right), \\
& \hat{\boldsymbol{y}}=\sqrt{\frac{2 \pi}{3}} i\left(\boldsymbol{Y}_{11}+\boldsymbol{Y}_{1,-1}+\boldsymbol{\Psi}_{11}+\boldsymbol{\Psi}_{1,-1}\right), \\
& \hat{\boldsymbol{z}}=\sqrt{\frac{4 \pi}{3}}\left(\boldsymbol{Y}_{10}+\boldsymbol{\Psi}_{10}\right) .
\end{align}

Because of boundary conditions, the magnetic fields sourced by DM will have a vanishing component along the radial direction. In particular, the ALP and DP signals will depend only on $\boldsymbol{\Phi}_{\ell m}$. In Table~\ref{tab:Philm} we tabulate $\boldsymbol{\Phi}_{\ell m}$ for $\ell \leq 3$. By choosing the $\hat{\boldsymbol{z}}$ direction to be aligned with the Earth's axis of rotation, $\boldsymbol{\Phi}_{\ell m}$ depends on the longitude $\phi$, which is zero at the Greenwich meridian and is positive due east, and the latitude $\pi/2-\theta$, which is $90^\circ$ at the North Pole and $0^\circ$ at the Equator.

\subsection{Signal}\label{app:signal}

The expected signal is a magnetic flux through the area subtended by the ion states in the trap. Therefore, the precise magnitude and orientation of the magnetic field is critical for an accurate estimate of the ion interferometer sensitivity. We discuss here the basics of the calculation of the magnetic field, following the thorough analysis in Refs.~\cite{Fedderke:2021aqo,Arza:2021ekq}.

Earth's crust has a DC conductivity that varies over the range $\sigma \sim 10^{-8}\,\text{eV} \div 10^{-6}\,\text{eV}$. This corresponds to a skin depth that varies between $\mathcal{O}(10)\,\text{km}\div \mathcal{O}(10^3)\,\text{km}$ for the frequency range we are most interested in, corresponding to masses $\in [10^{-17},10^{-13}]\,\text{eV}$. Given that the crust is only tens of km thick, this means that it does not behave as a good conductor for much of the parameter space of interest. However, Earth's mantle, particularly the lower mantle that begins at a depth of about $500\,\text{km}$, has a DC conductivity of $\sigma \sim 10^{-3}\,\text{eV}$, and is hundreds of km deep. Therefore, the lower mantle is, to a very good approximation, a conductive boundary for EM radiation in the frequency range of interest. Meanwhile, the plasma in the ionosphere and interplanetary medium act as an upper boundary for sufficiently small DM masses, turning Earth into an effective cavity with concentric spherical boundaries.

The signal magnetic field can be computed by solving Maxwell's equations in the presence of the DM effective current subject to the appropriate boundary conditions. As discussed in the main text, the effective currents for dark photon and ALP dark matter are
\begin{align}
    \boldsymbol{j}'_{\rm eff} &= - \epsilon m_{A'}^2 \boldsymbol{A'} \ , \\
    \boldsymbol{j}^a_{\rm eff} &= - g_{a\gamma\gamma} \partial_t a \boldsymbol{B}_\oplus \ .
\end{align}
The boundary condition problem can be solved by decomposing the EM fields in the effective Earth cavity in terms of eigenmodes of a cavity in the absence of any currents. These are further divided into so-called Transverse Electric and Magnetic modes, and given in terms of vector spherical harmonics $\boldsymbol{Y}_{\ell m} = Y_{\ell m} \hat{\boldsymbol{r}},~\boldsymbol{\Psi}_{\ell m} = r \nabla Y_{\ell m},~ \boldsymbol{\Phi}_{\ell m} = \boldsymbol{r}\times\nabla Y_{\ell m}$, where $Y_{\ell m}$ are scalar spherical harmonics, which are functions of polar and azimuthal angles, i.e., $\Omega=(\theta, \phi)$, and $ -\ell \leq m \leq \ell$ for $\ell \geq 0$. Maxwell's equations can be solved by writing them as second-order differential equations for $\boldsymbol{E},~\boldsymbol{B}$, 
\begin{align}
    (\nabla^2 - \partial_t^2) \boldsymbol{E} &= \partial_t\boldsymbol{j}_{\rm eff} \ , \\
    (\nabla^2 - \partial_t^2) \boldsymbol{B} &= - \nabla \times \boldsymbol{j}_{\rm eff} \ .
\end{align}
In the above equation of motion for the magnetic field, correctly accounting for the boundary conditions can be achieved by recalling that $\boldsymbol{j}_{\rm eff}$ will lead to a surface current on the conductive boundaries. 
We see from these equations that the electric field signal is sub-dominant for $m_{_{\rm DM}} R_\oplus \ll 1$, as the characteristic scale of a spatial derivative is $\nabla \gtrsim 1/R_\oplus$ due to the boundary conditions (see, e.g.,~\cite{Ouellet:2018nfr}), while the characteristic scale of the time derivative is $\partial_t \sim m_{_{\rm DM}}$. This yields the typical scaling of $\boldsymbol{B} \sim R_\oplus \boldsymbol{j}_{\rm eff}$ as argued in the main text.

The equations above were explicitly solved in the case of dark photon dark matter in Ref.~\cite{Fedderke:2021aqo}. To find the solution, an expression for the dark photon vector field in terms of vector spherical harmonics is required. Without loss of generality, we define the $\hat{\boldsymbol{z}}$  direction to be aligned with the Earth's axis of rotation, so that $\pi/2-\theta$ and $\phi$ correspond to latitude and longitude on Earth, respectively. In this situation, using the conventions of Appendix~\ref{sec:VSH}, one can write the dark photon field in a frame co-rotating with Earth at a frequency $\nu_d = 1/\text{day}$ as 
\begin{align}
    \boldsymbol{A'} = \sqrt{\frac{4\pi}{3}} \sum_{m=-1}^1 A'_m \left(\boldsymbol{Y}_{1m}+ \boldsymbol{\Psi}_{1m} \right)e^{- 2\pi i m \nu_d t}\ .
\end{align}
where $A'_{0}=\mathbb{R}[A'_z]$ and $A'_{\pm 1}=(\pm \mathbb{R}[A'_x] + i \mathbb{R} [A'_y])/\sqrt{2}$, with $A'_x$, $A'_y$ and  $A'_z$ the complex components of the dark photon field in Cartesian coordinates, e.g., $A'_x \propto \exp \left (im_{_{A'}}t-i\mathbf{k\cdot x} + i\phi_x \right )$. Note that the Earth's rotation only affects the $m=\pm 1$ contributions, since $\boldsymbol{Y}_{1m},\boldsymbol{\Psi}_{1m} \propto e^{i m\phi}$.
Using this decomposition, Ref.~\cite{Fedderke:2021aqo} found that the signal field for the Earth-ionosphere cavity in the limit $m_{A'}R_\oplus \ll 1$ is given by
\begin{align}
    \boldsymbol{B}_{_{A'}} = 
    \sqrt{\frac{\pi}{3}}\epsilon m_{A'}^2 R_\oplus \sum_{m=-1}^1 A'_m \boldsymbol{\Phi}_{1m} e^{- 2\pi i m \nu_d t} 
    \ ,
\end{align}
in the co-rotating frame.
In this expression, the vector spherical harmonic $\boldsymbol{\Phi}_{1m}$ is specified in the co-rotating frame so that it corresponds to the magnetic field as seen at a fixed location on the Earth's surface. In practice, we only consider masses significantly greater than $\nu_d$, so its importance is reduced. 

The case of ALP dark matter is more complicated, as the effective current is dependent not only on the axion field $a$, but also on the background magnetic field $\boldsymbol{B}_\oplus$. Earth's magnetic field is approximately dipole-like, but nevertheless contains significant overlap with higher multipoles. Therefore, a multipole expansion in terms of vector spherical harmonics is required to capture appropriately the contributions to the signal of different components of the background field. The International Geomagnetic Reference Field (IGRF)~\cite{Alken2021} is a commonly-agreed model for Earth's magnetic field, which is given in terms of spherical harmonic coefficients (also called Gauss coefficients) $g_{\ell m},~h_{\ell m}$ which have the dimensions of a magnetic field. The magnetic field is given as the gradient of a scalar potential, $\boldsymbol{B}_{\oplus} = - \nabla V$, where the potential is~\cite{Alken2021}
\begin{equation}
\begin{aligned}
    V(r,\theta,\varphi) &= R_\oplus \sum_\ell \sum_{m=0}^\ell \left(\frac{R_\oplus}{r} \right)^{\ell+1} \\
    & \qquad \times \left(g_{\ell m} \cos m \varphi + h_{\ell m} \sin m \varphi \right) \\
    & \qquad \times \frac{P_{\ell m}(\cos \theta)}{\sqrt{2l+1}} \ ,
\end{aligned}
\end{equation}
with $P_{\ell m}(\cos\theta)$ the usual associated Legendre polynomials and $r$ the radial distance from the centre of the Earth. This expression can be rewritten in terms of spherical harmonics $Y_{\ell m}$ with coefficients that have dimensions of a magnetic field, $C_{\ell m}$ as~\cite{Arza:2021ekq}
\begin{equation}
\begin{aligned}
    V(r,\theta,\varphi) &= R_\oplus \sum_\ell \sum_{m=-\ell}^\ell \left(\frac{R_\oplus}{r} \right)^{\ell+1} \\
    & \qquad \qquad \qquad \qquad \times C_{\ell m} Y_{\ell m}(\theta,\varphi) \ , \\
     C_{\ell m} &= (-1)^m\sqrt{\frac{4\pi(2-\delta_{0 m})}{2l+1}}\frac{g_{\ell m} - i h_{\ell m}}{2} \ .
    \label{eq:IGRFpotSphHarm}
\end{aligned}
\end{equation}
Numerically, the values of the $C_{\ell m}$ for the lowest-lying $\ell, m$ are given in Table~\ref{tab:clms}. The convention for these coefficients is fixed so that $C_{\ell,-m} = (-1)^m C^*_{\ell m}$.

Using the notation described above, the signal magnetic field resulting from solving Maxwell's equations in the presence of the axion effective current for the Earth-ionosphere cavity was obtained in Ref.~\cite{Arza:2021ekq}, where it was found that
\begin{equation}
\begin{aligned}\label{eq:BALP}
\boldsymbol{B}_{a} &= \mathbb{R}\left[- i g_{a\gamma\gamma} a \right] (m_a R_\oplus) \\
& \qquad \qquad \times \sum_{\ell}\sum_{m=-\ell}^\ell \frac{C_{\ell m}}{\ell} \left(\frac{R_\oplus}{r} \right)^{\ell+1} \boldsymbol{\Phi}_{\ell m} \, , 
\end{aligned}
\end{equation}
where $a \propto \exp \left (im_a t-i\mathbf{k\cdot x} + i \phi \right )$ is the ALP field.
Note that this field is \emph{not} affected by the daily rotation of the Earth, as the orientation of the signal field is fixed by the geomagnetic field. This is in contrast with the dark photon case, where the signal is set by the orientation of the dark photon itself. We will discuss the orientation of the signal as seen by the detector in Appendix~\ref{supp:signal} below.


\begin{table*}[t]
    \centering
    \renewcommand{\arraystretch}{1.5}
    \begin{tabular}{c|c|c|c|c}
    \hline
    \hline
    \multicolumn{5}{c}{$C_{\ell m}$ [nT]} \\
    \hline
         $~~\ell~~$ & $m=0$ & $m=1$ & $m=2$ & $m=3$ \\
         \hline
         $1$ & $-30061.55$ & $2046.20 + 6545.70\,i$ & -- & -- \\
         $2$ & $-2024.94$ & $-3303.58-3522.85\,i$ & $1868.14+949.04\,i$ & -- \\
         $3$ & $920.61$ & $2283.94 - 49.36\,i$ & $1185.88 - 223.97\,i$ & $-441.21-512.46\,i$ \\
        \hline
        \hline
    \end{tabular}
    \caption{Numerical values in $\text{nT}$ for the predicted 2025 IGRF spherical harmonic coefficients $C_{\ell m}$ defined in Eq.~\eqref{eq:IGRFpotSphHarm} in terms of the usual Gauss coefficients $g_{\ell m},~h_{\ell m}$ provided in Ref.~\cite{Alken2021}.}
    \label{tab:clms}
\end{table*}

\subsection{Shielding around the ion trap }
\label{sec:shielding}

Real conductive shields are not perfect, resulting in residual magnetic and electric fields penetrating into the shielded regions. This is of particular importance for the dark matter mass ranges we consider here, as these correspond to EM fields oscillating at very low frequencies, which are more difficult to shield against. The crucial parameters that define whether a shield is effective or not are the skin depth, $\delta_s$, and the thickness of the shield, $\Delta$. For an EM wave with $\omega\epsilon_p \ll \sigma_0$, where $\epsilon_p$  is the relative permittivity of the material with DC conductivity $\sigma_0$, the skin depth is given by
\begin{align}
    \delta_s = \sqrt{\frac{2}{\omega \mu \sigma_0}} \ ,
\end{align}
where $\mu$ is the magnetic permeability. In the same limit, the real and imaginary parts of the EM wavevector are identical, and are $k_{r,i} = 1/\delta_s$. We can therefore define the propagation constant as $\gamma = ik = (1+i)/\delta_s$ such that any plane-wave propagation factor is $e^{\gamma x}$.

\begin{figure}
    \centering
    \includegraphics[width=0.48\textwidth]{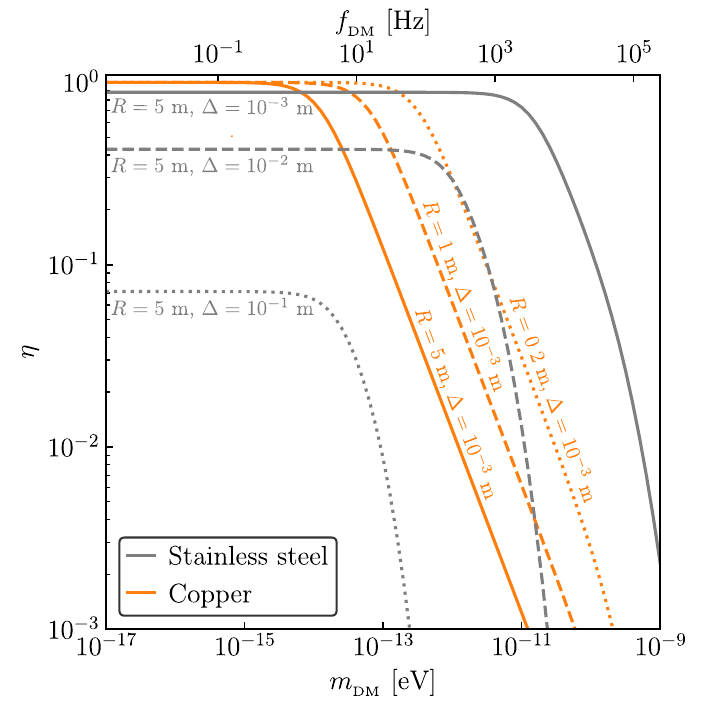}
    \caption{ Shielding efficiency $\eta$ as defined in Appendix~\ref{sec:shielding} for various possible shields
    . We assume a spherical shield made of either copper ($\mu = 1,~\sigma = 2.55\times10^7$~S/m) or stainless steel ($\mu = 10^3,~\sigma = 2\times10^4$~S/m), and take various configurations for the typical size $R$ and the thickness of the shield $\Delta$.}
    \label{sfig:efficiency}
\end{figure}

As an approximation, we treat the possible magnetic shield around the ion trap as  a spherical shell of radius $R$ and thickness $\Delta$. This enables us to find a closed form for the shielding efficiency, defined as
\begin{align}
    \eta \equiv \frac{|\boldsymbol{B}_{\rm shield}|}{|\boldsymbol{B}_{\rm naive}|} \ ,
\end{align}
which gives the ratio of the magnitude of the magnetic field inside the shield relative to its value in the absence of a shield. We further assume that the applied magnetic field outside the shield is uniform, so that the magnetic field inside is also uniform, and $\eta$ is independent of position inside the shield. Since the characteristic scale of the signal magnetic field is much larger than the apparatus or the shield size, the uniform-field approximation is justified. In this simplified setup, the shielding efficiency is known fully analytically~\cite{HoburgShield}, and is given by
\begin{widetext}
\begin{align}
    \eta^{-1} &= \Bigg\vert \frac{1}{3 R^3 \gamma^3 \mu} \left( f_s(R,\Delta,\gamma,\mu) \sinh(\gamma\Delta) + f_c(R,\Delta,\gamma,\mu) \cosh(\gamma\Delta) \right) \Bigg\vert \ ,
\end{align}
with 
\begin{align}
    f_s(R,\Delta,\gamma,\mu) &= (R\gamma)^4 - 2 R^3 \Delta \gamma^4+ (R\gamma)^2(1+2\mu^2+\gamma^2\Delta^2) + R\Delta \gamma^2(\mu(3-2\mu)-1) \nonumber \\ & \qquad -(\mu-1)(1+ 2\mu +\gamma^2\Delta^2 ) \ , \\
    f_c(R,\Delta,\gamma,\mu) &=3\mu (R\gamma)^3 + R^2\Delta\gamma^3 (1 - 4\mu ) + R\Delta^2\gamma^3(\mu-1) + \gamma\Delta(2\mu^2-\mu-1) \ .
\end{align}
\end{widetext}
We use this unintuitive but fully general expression to produce the modulus of the shielding factors shown in Fig.~\ref{sfig:efficiency}, and to represent in Fig.~\ref{fig:reach} the effects on our expected sensitivity of two different shield choices.

The most useful limits for this shielding efficiency expression are those in which $\gamma\Delta \to 0$ and $\gamma\Delta \to \infty$, corresponding to the zero-frequency and high-frequency regimes. In these limits, and further taking the thin-wall approximation in which $\Delta \to 0$, we find
\begin{align}
    \eta_{\gamma\Delta \to 0,\Delta\to 0} &\simeq 1- \frac{2\Delta}{3R}\frac{(\mu-1)^2}{\mu} \ , \\
    \eta_{\gamma\Delta \to \infty,\Delta \to 0} &\simeq \Bigg\vert \Bigg[\cosh(\gamma\Delta) \nonumber \\ & + \gamma\Delta\left( \frac{R}{3\Delta \mu}- \frac{2}{3\mu}\right) \sinh(\gamma\Delta) \Bigg]^{-1} \Bigg\vert \ .
\end{align}
The first limit is valid as long as $\mu\ll R/\Delta$. These expressions reproduce correctly the small- and large-mass shielding efficiencies shown in Fig.~\ref{sfig:efficiency}.

\section{Signal Analysis}
\label{supp:signal}

In this appendix, we provide additional details regarding the analysis of time-dependent signals in trapped-ion interferometers. In particular, we show how to perform the analysis in frequency space, and how to take into account finite-sampling effects.

\subsection{Conventions}

We briefly summarize here our conventions for working in the frequency domain. We adopt the following convention for the Fourier transforms (FT) of a function $f(t)$, 
\begin{equation}
\widetilde{f}(\omega)=\int_{-\infty}^{\infty} \mathrm{d} t e^{i \omega t} f(t) \, , \quad f(t)=\int_{-\infty}^{\infty} \frac{\mathrm{d} \omega}{2 \pi} e^{-i \omega t} \widetilde{f}(\omega) \, .
\end{equation}
The power of $f$ is defined as
\begin{equation}
P_f = \lim_{T\rightarrow\infty} \frac{1}{2T}\int_{-T}^{T} \mathrm{d}t \, \langle |f(t)|^2\rangle = \int_{-\infty}^{\infty} \frac{\mathrm{d}\omega}{2\pi}S_f(\omega)\, ,
\end{equation}
where $S_f$ is the power spectral density (PSD) of $f$. From these definitions, it immediately follows
that
\begin{equation}
2\pi \delta (\omega-\omega')S_f(\omega) = \langle \widetilde{f}(\omega) \widetilde{f}^*(\omega')\rangle \, .
\end{equation}

\subsection{Phase shift signal}

Consider an interferometric sequence initiated at time $t$. As explained in the main text, the phase shift is measured at time $t+\Delta t$. In the time domain, the phase shift induced by a uniform external magnetic field can be expressed as 
\begin{equation}\label{supp:eq:phase_shift}
\begin{aligned}
    \Delta \Phi(t) &\simeq 2e \int_t^{t+\Delta t}
    \boldsymbol{B} \cdot\frac{\mathrm{d}\boldsymbol{\Sigma}}{\mathrm{d}t'} \mathrm{d}t' 
    \\
    &= 2e \int_{-\infty}^{\infty} W(t',t,\Delta t) \,  \boldsymbol{B} \cdot\frac{\mathrm{d}\boldsymbol{\Sigma}}{\mathrm{d}t'} \mathrm{d}t' \, , 
\end{aligned}
\end{equation}
where the window function is defined in terms of Heaviside functions as $W(t',t,\Delta t) = \Theta(t'-t)\Theta(t+\Delta t-t')$. The last line of Eq.~\eqref{supp:eq:phase_shift} is a convolution of the window function with the projection of the magnetic field component along $\hat{\boldsymbol{\Sigma}}$. Therefore, by the convolution theorem, the Fourier transform (FT) of Eq.~\eqref{supp:eq:phase_shift} is given by 
\begin{equation}
\widetilde{\Delta  \Phi}(\omega) = 2e \, \widetilde{W}(\omega, \Delta t)  \widetilde{\boldsymbol{B}}(\omega) \cdot \frac{\mathrm{d}\boldsymbol{\Sigma}}{\mathrm{d}t} \, ,
\end{equation}
where $\widetilde{W}(\omega, \Delta t) = i\left (1-e^{i\omega \Delta t} \right)/\omega$.
Equipped with these expressions, the power spectral density (PSD) of the measured phase shift, which is related to the FT of the measured phase by $\langle \Delta \widetilde{\Phi}^*(\omega')\Delta \widetilde{\Phi}(\omega) \rangle \equiv 2\pi S_{\Delta \Phi}(\omega)\delta(\omega-\omega')$, can be expressed as
\begin{equation} \label{supp:eq:PSD}
\begin{aligned}
S_{\Delta \Phi}(\omega) &= 8\pi e^2 \, \left < \left | \widetilde{\boldsymbol{B}}(\omega) \cdot  \frac{\mathrm{d}\boldsymbol{\Sigma}}{\mathrm{d}t} \right |^2 \right > \left | \widetilde{W}(\omega, \Delta t) \right |^2 \, \\
& = 8\pi e^2 \, \left < \left | \widetilde{\boldsymbol{B}}(\omega) \cdot  \frac{\mathrm{d}\boldsymbol{\Sigma}}{\mathrm{d}t} \right |^2 \right > (\Delta t)^2 \, \mathrm{sinc}^2\left (\frac{\omega \Delta t}{2} \right) \\
& = 2\pi \left < \left | \widetilde{\boldsymbol{B}}(\omega) \cdot \hat{\boldsymbol{\Sigma}} \right |^2 \right > \left (\frac{e N k_{\rm eff} y_d \Delta t}{m_{\rm ion}} \right )^2 \, \mathcal{T}(\omega) \, ,
\end{aligned}
\end{equation}
where $\mathrm{sinc}(x) = \sin(x)/x$, $\mathcal{T}(\omega) = \mathrm{sinc}^2(\omega \Delta t/2)$ is the detector transfer function and in the final line we used the definition of the rate of change of the area subtended by the wavepackets, i.e. $\mathrm{d}\boldsymbol{\Sigma}/\mathrm{d}t = \hat{\boldsymbol{\Sigma}} \, N k_{\rm eff} y_d/2m_{\rm ion} $. In light of this transfer function, trapped-ion interferometers effectively act as low-pass filters for $\omega \lesssim 1/T$. Indeed, for $\omega \lesssim 1/\Delta t$, the PSD grows quadratically with the interrogation time $T$; for $\omega \gtrsim 1/\Delta t$, instead, the PSD is independent of $\Delta t$ and $1/\omega^2$ suppressed.

\subsection{Dark matter signal}

In light of the dependence of the signal on the projection of the magnetic field onto $\hat{\boldsymbol{\Sigma}}$, the DM signal can be maximized if the direction of the magnetic field is known {\it a priori}. In the DP case, the direction of the DM-sourced magnetic field depends on the polarization of the field. Assuming the standard scenario where the spin-1 particle is unpolarized, the polarization of the DP varies every coherence patch, and, crucially, is not known from the outset. Furthermore, in light of the independence of the DM-sourced magnetic field on $\Sigma$, the expectation value of the signal is insensitive to the location of the experiment on the Earth's surface.
This is in stark contrast with the ALP case: because the direction of the ALP DM-sourced magnetic field is set by the Earth's magnetic field, it is possible to orient the trap so as to be maximally sensitive to the DM signal. Furthermore, in light of the dependence of the ALP-sourced magnetic field on latitude and longitude, the size of the expected signal can be maximized by placing the trapped-ion interferometer at specific locations.

\begin{figure*}
    \centering
    \includegraphics[width=0.95\textwidth]{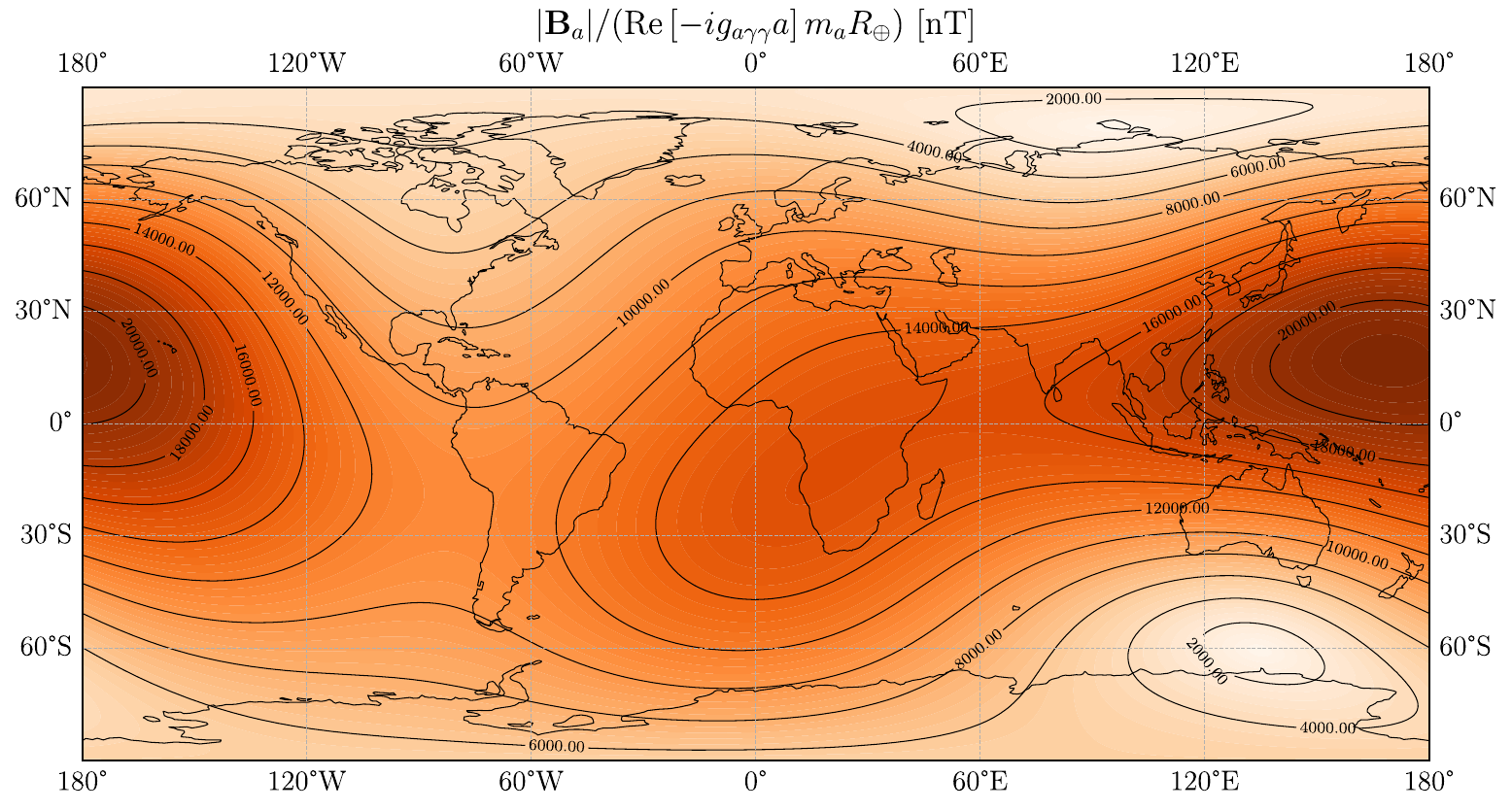}
    \caption{Normalized magnitude of the magnetic field sourced by ALP dark matter, $\boldsymbol{B_a}$, in nT at the Earth's surface. This is equivalent to taking $\mathbb{R}\left[- i g_{a\gamma\gamma} a \right] (m_a R_\oplus)=1$ and $r=R_\oplus$ in Eq.~\eqref{eq:BALP}.} 
    \label{sfig:BALP_map}
\end{figure*}

In light of these observations, we assume two different orientations for ALP and DP searches. In the former case, we assume a trap with $\hat{\boldsymbol{\Sigma}}\parallel \boldsymbol{B_a}$. The ALP-sourced magnetic field strength is maximal at approximately $(169^\circ \mathrm{W},18^\circ \mathrm{S})$, close to the Bikini Atoll in the Pacific Ocean, as shown in Fig.~\ref{sfig:BALP_map}. At this location, $B_a/(g_{a\gamma\gamma} a m_a R_\oplus) \approx 0.22$~G. 
Close to the geomagnetic poles, $B_a/(g_{a\gamma\gamma} a m_a R_\oplus )\approx 0$~G. Note that this does not imply that the Earth's magnetic field is zero at the geomagnetic poles, where instead it is maximal and $\gtrsim 0.6$~G. Simply, it states that the component of the magnetic field sourced by the ALP in the $\boldsymbol{\Phi}_{\ell m}$ direction is approximately zero. In our projection, we assume that the sensor is located in Los Angeles, where we anticipate $B_a/(g_{a\gamma\gamma} a m_a R_\oplus) \approx 0.11$~G. This geographical choice degrades the amplitude of the signal by a factor of two, a correction which, however, is still well within the error bars of our noise model. For DP searches, without loss of generality, we assume a sensor with $\hat{\boldsymbol{\Sigma}} \parallel \hat{\boldsymbol{\phi}}$. 

For our choice of sensor orientation, one can compute the expected PSD of the DM signal for a given DM mass and coupling to photons. For the ALP case, it trivially follows that the DM-dependent part of the PSD in Eq.~\eqref{supp:eq:PSD} is given by 
\begin{widetext}
\begin{equation}
\begin{aligned}
\left < \left | \widetilde{\boldsymbol{B}}_a(\omega) \cdot \hat{\boldsymbol{\Sigma}} \right |^2 \right > = 
\left < \left | \widetilde{\boldsymbol{B}}_a(\omega) \right |^2 \right > & \approx g_{a\gamma\gamma}^2 \left \langle |a|^2 \right \rangle  (m_a R_\oplus)^2 (0.11~\text{G})^2 \\
 & = 2 g_{a\gamma\gamma}^2  R_\oplus^2 \rho_{_\mathrm{DM}} (0.11~\text{G})^2 \, ,
\end{aligned}
\end{equation}
where we used the definitions of $\boldsymbol{\Phi}_{\ell m}$ provided in Table~\ref{tab:Philm}, assumed a sensor located in Los Angeles and employed the statistical properties of the ALP field, namely $\langle |a|^2 \rangle = 2\rho_{_\mathrm{DM}}/m_a^2 $. For the DP case, this same quantity is given by
\begin{equation}
\begin{aligned}
\left < \left | \widetilde{\boldsymbol{B}}_{_{A'}}(\omega) \cdot \hat{\boldsymbol{\Sigma}} \right |^2 \right > & = \frac{1}{4} \epsilon^2 m_{A'}^4 R_\oplus^2 \left [ \left \langle |A'_0|^2 \right \rangle \sin^2 \theta + \frac{1}{2} \left \langle |A'_1|^2 \right \rangle \cos^2 \theta + \frac{1}{2} \left \langle |A'_{-1}|^2 \right \rangle \cos^2 \theta\right ]  \\
& = \frac{1}{6} \epsilon^2 m_{A'}^2 R_\oplus^2 \rho_{_\mathrm{DM}} \, ,
\end{aligned}
\end{equation}
\end{widetext}
where we used the statistical properties of unpolarized spin-1 ultralight DM, namely $\langle A'_m A'^*_{m'}\rangle = \langle |A'_m|^2 \rangle \delta_{mm'}$ and $\langle |A'_m|^2 \rangle = 2\rho_{_\mathrm{DM}}/3 m_{_{A'}}$. Note that, as anticipated, the expected signal as measured by the sensor is independent of longitude and latitude.  

\subsection{Likelihood and test statistics}

In light of the stochasticity of the DM signal and of all dominant noise sources (e.g., shot-noise and ambient magnetic noise), the PSD of the signal and noise will be random variables. It is therefore advantageous to perform the analysis in the frequency domain.  In this work, we assume that the measurement campaign lasts for an integration time $T_\mathrm{int}$. Therefore, the smallest resolvable angular frequency is $\Delta \omega = 2\pi/T_\mathrm{int}$. Because the amplitude of the DM field and of ambient magnetic noise are Rayleigh distributed, the discrete signal and noise PSDs, $S_s^{(k)}$ and $S_n^{(k)}$, are exponentially distributed. Consequently, for a data stream $d(t)=s(t)+n(t)$, where $s(t)$ and $n(t)$ are the DM signal and noise, respectively, the likelihood for a model $\mathcal{M}$ with parameter vector $\boldsymbol{\theta}$ takes the form
\begin{equation}
\mathcal{L}(d | \mathcal{M},\boldsymbol{\theta}) = \prod_{k=1}^{N-1} \frac{\exp \left (-\frac{|\widetilde{d}^{(k)}|^2}{S_s^{(k)}+S_n^{(k)}}\right )}{\pi \left (S_s^{(k)}+S_n^{(k)} \right )} \, .
\end{equation}
Here, $\widetilde{d}^{(k)}$ is the $k$-th component of the discrete FT of the data centered at $\omega_k = 2\pi k /T_\mathrm{int}$ and $\boldsymbol{\theta} = \{\boldsymbol{\theta}_{s},\boldsymbol{\theta}_{n}\}$, where $\boldsymbol{\theta}_{s}$ and $\boldsymbol{\theta}_{n}$ are signal (e.g., DM mass and coupling to photons) and nuisance parameter (e.g., the magnitude of the ambient magnetic field) vectors, respectively. Note that $T_\mathrm{int} = (N - 1)\Delta t$, where we assume negligible dead time between successive experimental runs and $k = 0$ is omitted from the product, since it is degenerate with a tower of static effects.

To set upper limits on the coupling $\kappa^2 \in \{g_{a\gamma\gamma}^2, \epsilon^2 \}$ we use the test statistic
\begin{equation}
\begin{aligned}
q(m_{_\mathrm{DM}},\kappa^2) & = -2 \ln \left ( \frac{\mathcal{L}(d | \mathcal{M},\{m_{_\mathrm{DM}},\kappa\},\hat{\hat{\boldsymbol{\theta}}}_n)}{\mathcal{L}(d | \mathcal{M},\{m_{_\mathrm{DM}},\widehat{\kappa^2}
\},\hat{\boldsymbol{\theta}}_n)} \right ) \\
& \qquad \times \Theta \left (\kappa^2-\widehat{\kappa^2} \right ) \, ,
\end{aligned}
\end{equation}
where $\widehat{\kappa^2}$ and $\hat{\boldsymbol{\theta}}_n$ are, respectively, the square of the DM-photon coupling and the nuisance parameter vector that maximize the likelihood at a fixed DM mass (i.e., unconditional maximum-likelihood estimators) and $\hat{\hat{\boldsymbol{\theta}}}_n$ is the nuisance parameter vector that maximizes the likelihood for fixed $\kappa^2$ and DM mass (i.e., conditional maximum-likelihood estimator).

In the regime $T_\mathrm{int} \gg \tau_c$, the experiment is able to resolve the spectral content of the DM signal. By applying Wilks' Theorem~\cite{Wilks:1938dza} it therefore follows that $q(m_{_\mathrm{DM}},\kappa^2)$ at fixed $m_{_\mathrm{DM}}$ is a half chi-squared distribution with one degree of freedom, such that the 95\% confidence limits on $\kappa^2$ are set by $q_{95\%}\approx-2.71$~\cite{Cowan:2010js}. To set upper limits on $\kappa$, we employ the Asimov approach, i.e., $|\widetilde{d}^{(k)}|^2 \rightarrow S_n^{(k)}$ so that $\widehat{\kappa^2} = 0$, which has been used in several works on ULDM detection (e.g., Refs.~\cite{Foster:2017hbq}). Furthermore, we assume $\hat{\boldsymbol{\theta}}_n \simeq \hat{\hat{\boldsymbol{\theta}}}_n$. In this regime, the signal is spread over multiple frequency bins; because of power conservation, $S_s^{(k)} \ll S_n^{(k)}$ for all $k$, such that the test statistic takes the form
\begin{equation}
\begin{aligned}
q(m_{_\mathrm{DM}},\kappa^2) & \approx \sum_{k}  \left (\frac{S_s^{(k)}}{S_n^{(k)}} \right )^2 \\ & \simeq \frac{T_\mathrm{int}}{2\pi} \int_0^\infty \mathrm{d}\omega \left (\frac{S_s(\omega)}{S_n(\omega)} \right )^2 = \mathrm{SNR}^2 \, , 
\end{aligned}
\end{equation}
where the second line is valid in the continuum limit and corresponds to the definition of the signal-to-noise ratio (SNR) squared. The DM signal has support over a frequency range set by the DM speed distribution, i.e. $\Delta \omega_{_\mathrm{DM}} \simeq 2\pi/\tau_c$. Approximating the speed distribution of the Standard Halo Model to a box function and by imposing power conservation, we find
\begin{equation}
S_s(\omega) \simeq \frac{2 \pi P_{_\mathrm{DM}}}{\Delta \omega_{_\mathrm{DM}}} \Theta\left(\omega-m_{_\mathrm{DM}}\right) \Theta\left(\Delta \omega_{_\mathrm{DM}}+ m_{_\mathrm{DM}}-\omega\right) \, ,
\end{equation}
where $P_{_\mathrm{DM}}$ is the power of the DM signal.
With these approximations and for a background that is approximately constant between $m_{_\mathrm{DM}}$ and $m_{_\mathrm{DM}}+\Delta \omega_{_\mathrm{DM}} $, the SNR becomes
\begin{equation}\label{eq:SNR_approx}
\mathrm{SNR} \simeq \frac{P_{_\mathrm{DM}}}{S_n(m_{_\mathrm{DM}})} \sqrt{T_\mathrm{int}\tau_c} \, .
\end{equation}
Since $P_{_\mathrm{DM}} = \langle|\Delta \Phi_{_\mathrm{DM}}|^2\rangle \propto \kappa^2$, where $\langle|\Delta \Phi_{_\mathrm{DM}}|^2\rangle$ is the amplitude squared of the time-averaged DM-induced phase shift, the 95\% upper limit on $\kappa$ scales with $(T_\mathrm{int}\tau_c)^{-1/4}$ and is determined by solving Eq.~\eqref{eq:SNR_approx} for $\mathrm{SNR}\approx 1.64$. 

In the regime $T_\mathrm{int} \ll \tau_c$, the experiment is no longer able to resolve the spectral content of the DM signal. Because the signal would be contained within a single frequency bin of size $\Delta \omega$, Wilks' Theorem no longer applies. As shown in Ref.~\cite{Berlin:2020vrk}, the test statistic takes the approximate form
\begin{equation}
q(m_{_\mathrm{DM}},\kappa^2) \approx 2 \ln \left ( \frac{S_s^{(k)}}{S_n^{(k)}} \right ) - 2 \, ,
\end{equation}
for $k$ such that  $m_{_\mathrm{DM}} \in [\omega_k-\Delta \omega/2, \omega_k+\Delta \omega/2 ]$, and the the median expected 95\% limit on $\kappa^2$ corresponds to $q_{95\%} \approx 3.05$. By imposing power conservation, $S_n^{(k)} = S_n(\omega_k)$ for a background that does not vary appreciably over $\Delta \omega$, and for the DM signal we find
\begin{equation}
S_s^{(k)} = \frac{1}{\Delta \omega}\int_{\omega_k-\Delta \omega/2}^{\omega_k+\Delta \omega/2} \mathrm{d} \omega \, S_s(\omega) \simeq  \frac{2 \pi P_{_\mathrm{DM}}}{\Delta \omega} \, .
\end{equation}
Consequently, the SNR is given by
\begin{equation}\label{eq:SNR_short}
\mathrm{SNR} \simeq \frac{P_{_\mathrm{DM}}}{S_n(m_{_\mathrm{DM}})} T_\mathrm{int} \, ,
\end{equation}
the 95\% upper limit on $\kappa$ scales with $(T_\mathrm{int})^{-1/2}$ and is determined by solving Eq.~\eqref{eq:SNR_short} for $\mathrm{SNR}\approx 12.5$. 

\subsection{Finite sampling effects}

The experimental sampling rate $f_\star$ is comparable to a subset of the signal frequencies that we wish to measure. Assuming zero dead time (i.e., successive phase shift measurements are temporally separated by $\Delta t$), $f_\star = 1/\Delta t$. For $\Delta t = 1$~s, DM signals with $f_{_\mathrm{DM}} \equiv m_{_\mathrm{DM}}/2\pi \gtrsim 1$~Hz would be affected by spectral distortions and aliasing. This can be illustrated by assuming a sampling angular frequency $\omega_\star=2\pi f_\star$, so that the sampled phase shift takes the form $\Delta \Phi_\mathrm{sampled} (t) = \Delta \Phi(t) \Sh(t)$, where $\Sh(t) = \sum_{k = -\infty}^{\infty} e^{i\omega_\star k t}$ is the Dirac comb. The FT of the sampled phase shift then takes the form
\begin{equation}
\widetilde{\Delta\Phi}_\mathrm{sampled}(\omega) = \widetilde{\Delta\Phi}(\omega) + \sum_{k\neq 0}^{\infty} \widetilde{\Delta\Phi}(\omega + k \omega_\star) \, ,
\end{equation}
so that the sampled phase shift's PSD takes the form
\begin{equation}
S_{{\Delta \Phi},\mathrm{sampled}}(\omega) = S_{\Delta \Phi}(\omega) + \sum_{k\neq 0}^{\infty} S_{\Delta \Phi}(\omega + k \omega_\star) \, ,
\end{equation}
Importantly, this expression suggests that any signal or noise component that lies above the experiment's sampling frequency would be identified at a lower frequency (i.e., aliasing) and added to the true 
PSD at lower frequencies (i.e., folding), thus giving rise to spectral distortions. 

Provided that the spectrum of colored noise at frequencies greater than 1~Hz is subdominant or undergoes a power-law fall-off, it may still be possible to detect a DM candidate with $m_\mathrm{DM} \gtrsim 10^{-15}$~eV, as extensively studied in Ref.~\cite{Badurina:2023wpk} in the context of ULDM detection with atom gradiometers and interferometers. Since the width of the spectral lineshape is set by the DM mass, aliased DM signals can be disentangled from non-aliased ones provided that $T_\mathrm{int}/N_\mathrm{stacks} \gtrsim \tau_c$, where $N_\mathrm{stacks}$ is the number of stacks (i.e., after data collection, one breaks up the time series into $N_\mathrm{stacks}$ portions of duration $T_\mathrm{int}/N_\mathrm{stacks}$). To identify correctly folded signals, it is sufficient for the experimental (stacked) frequency resolution to be approximately five times greater than the signal linewidth. Alternatively, it may be possible to avoid spectral distortions via unequal sampling~\cite{Eyer:1998jr,10.1111/j.1365-2966.2006.10762.x}.  We leave a more detailed analysis to future work. 

\bibliography{main}
\bibliographystyle{apsrev4-2}

\end{document}